\documentclass[11pt]{article}
\newif\iffigs\figstrue

\usepackage{amsbsy}
\usepackage[title]{appendix}
\usepackage{epsfig,latexsym}
\usepackage{amsmath}
\usepackage{verbatim}
\usepackage{mathrsfs}
\usepackage{amssymb}
\usepackage{ytableau}

\DeclareMathAlphabet{\mathpzc}{OT1}{pzc}{m}{it}

 \csname
@addtoreset\endcsname{equation}{section}

\def\gz0{\gamma^{0}}

\def\a{\alpha}

\def\n{\nu}

\def\bec{\begin{center}}
\def\ec{\end{center}}

\def\12{\frac{1}{2}}

\def\pr{\partial}

\def\DH{\rm I\kern-1.5pt\rm H\kern-1.5pt\rm I}

\def\DR{\rm I\kern-1.45pt\rm R}
\def\DC{\kern2pt {\hbox{\sqi I}}\kern-4.2pt\rm C}

\newcommand{\beq}{\begin{equation}}
\newcommand{\eeq}{\end{equation}}
\newcommand{\bea}{\begin{eqnarray}}
\newcommand{\eea}{\end{eqnarray}}
\newcommand{\bi}{\begin{itemize}}
\newcommand{\ei}{\end{itemize}}

\usepackage{slashed}

\usepackage{float}

\usepackage{caption}

\usepackage{epsf}
\usepackage{amsmath}

\definecolor{alizarin}{rgb}{0.82, 0.1, 0.26}

\usepackage{multirow}

\usepackage{epsfig}
\usepackage{cite}
\usepackage{color,colordvi}

\newcounter{hran}

\makeatletter
\renewcommand\section{\@startsection {section}{1}{\z@}%
                               {-3.5ex \@plus -1ex \@minus -.2ex}%
                               {2.3ex \@plus.2ex}%
                               {\normalfont\large\bfseries}}
\makeatother

\vspace{0.5cm}

\begin{document}
\thispagestyle{empty}

\vspace{15pt}

\begin{center}


{\Large\sc On Classical Stability with Broken Supersymmetry}\\


\vspace{50pt}
{\sc I.~Basile${}^{\; a}$, J.~Mourad${}^{\; b}$  \ and \ A.~Sagnotti${}^{\; a}$}\\[15pt]

{${}^a$\sl\small
Scuola Normale Superiore and INFN\\
Piazza dei Cavalieri, 7\\ 56126 Pisa \ ITALY \\
e-mail: {\small \it basile@sns.it, sagnotti@sns.it}}\vspace{8pt}

{${}^b$\sl\small APC, UMR 7164-CNRS, Universit\'e Paris Diderot -- Paris 7 \\
10 rue Alice Domon et L\'eonie Duquet \\75205 Paris Cedex 13 \ FRANCE
\\ }e-mail: {\small \it
mourad@apc.univ-paris7.fr}\vspace{10pt}

\vspace{16pt}

\vspace{25pt} {\sc\large Abstract}
\end{center}
We study the perturbative stability of four settings that arise in String Theory, when dilaton potentials accompany the breaking of Supersymmetry, in the tachyon--free $USp(32)$ and $U(32)$ orientifold models, and also in the heterotic $SO(16)\times SO(16)$ model. The first two settings are a family of $AdS_3 \times S^7$ vacua of the orientifold models and a family of $AdS_7 \times S^3$ vacua of the heterotic model, supported by form fluxes, with small world--sheet and string--loop corrections within wide ranges of parameters. In both cases we find some unstable scalar perturbations, as a result of mixings induced by fluxes, confirming for the first class of vacua a previous result. However, in the second class of vacua they only affect the $\ell=1$ modes, so that a $\mathbb{Z}_2$ projection induced by an overall parity in the internal space suffices to eliminate them, leading to perturbative stability. Moreover, the constant dilaton profiles of these vacua allow one to extend the analysis to generic potentials, thus exploring the possible effects of higher--order corrections, and we exhibit wide nearby regions of perturbative stability. The solutions in the third setting have nine--dimensional Poincar\'e symmetry. They include regions with large world--sheet or string--loop corrections, but we show that these vacua have no perturbative instabilities. Finally, the last setting concerns cosmological solutions in ten dimensions where the ``climbing'' phenomenon takes place: they have bounded string--loop corrections but large world--sheet ones close to the initial singularity. In this case we find that perturbations generally decay, but homogeneous tensor modes exhibit an interesting logarithmic growth that signals a breakdown of isotropy. If the Universe then proceeds to lower dimensions, milder potentials from other branes force all perturbations to remain bounded.
\noindent

\vfill
\noindent

\baselineskip=18pt

\newpage
\tableofcontents
\newpage

\setcounter{page}{2}
\setcounter{equation}{0}
\section{Introduction} \label{sec:intro}

When Supersymmetry is broken in String Theory~\cite{stringtheory}, one is inevitably confronted with stability issues. These are particularly severe when tachyons emerge, but even in the few cases when this does not occur runaway dilaton potentials destabilize the original ten--dimensional Minkowski vacua. Supergravity~\cite{sugra} is the key tool to analyze these systems, within its limits of applicability.

There are three ten--dimensional string models with broken Supersymmetry and no tachyons, and two of them are orientifolds~\cite{orientifold} of closed--string models. The first of these is the $USp(32)$ model~\cite{sugimoto} with ``Brane Supersymmetry Breaking'' (BSB)~\cite{bsb}, a peculiar type--IIB orientifold which combines the presence of a gravitino in the low--energy Supergravity with a non--linearly realized Supersymmetry~\cite{dmnonlinear}. The second has no Supersymmetry at all, and is the $U(32)$ $0'B$ model~\cite{u32}, a non--tachyonic orientifold of the tachyonic $0B$ model~\cite{0b}. In both cases, an exponential dilaton potential emerges at the (projective--)disk level, which reflects the residual tension of the branes and orientifolds that are present in ten dimensions. In the Einstein frame it reads
\beq
V \ = \ T \ e^{\,\frac{3}{2}\,\phi} \ . \label{pot32}
\eeq
There is finally a heterotic model of this type~\cite{so16xso16}, with an $SO(16) \times SO(16)$ gauge group. Its torus amplitude gives a contribution that in the string frame is independent of the dilaton, but in the Einstein frame this translates into the dilaton potential~\cite{review}
\beq
V \ = \ T \ e^{\,\frac{5}{2}\,\phi} \ . \label{pot52}
\eeq
In the first case there is a class of $AdS_3 \times S^7$ vacua~\cite{ms17}, which are sustained by three--form (electric) fluxes that permeate $AdS_3$, and where the dilaton has constant profiles. In the second case similar steps lead to a class of $AdS_7 \times S^3$ vacua~\cite{ms17}, which are sustained by three--form (magnetic) fluxes that permeate the internal $S^3$, and where the dilaton has again constant profiles~\footnote{There are also descriptions in terms of dual potentials, which involve seven--form fluxes in internal space for the orientifold models and in spacetime for the heterotic one.}. In both cases there are wide corners of parameter space where large radii and small string couplings are possible.

In this paper we investigate perturbative stability in the presence of broken Supersymmetry in four types of settings. The first two concern these $AdS$ vacua for the orientifold and heterotic models, where fluxes counteract the effects of gravity. The third concerns the Dudas--Mourad solutions with nine--dimensional Poincar\'e symmetry~\cite{dm_9Dsolution}, the first that were found for the low--energy equations of String Theory when the potentials \eqref{pot32} or \eqref{pot52} arise. No form fluxes are present in this case, and the resulting vacua include corners where the string coupling is large or curvature corrections are expected to be important. The fourth concerns cosmological solutions with nine--dimensional Euclidean symmetry, which were also first found in~\cite{dm_9Dsolution} and where the string coupling has an upper bound, so that string corrections are in principle under control, while curvature corrections are still large close to the initial singularity. Some generalizations of this solution were found later in~\cite{russo}, and were elaborated upon in detail in~\cite{climbing}, where they were associated to the picture of a ``climbing scalar''. Amusingly, the orientifold potential in eqs.~\eqref{pot32} corresponds indeed to a ``critical'' logarithmic slope where the scalar (the dilaton in ten dimensions or, in lower dimensions, a mixing of it with the breathing mode of the internal space~\cite{integrable}) becomes compelled to climb up the potential when emerging from the initial singularity. In the subsequent descent the potential energy thus collected can give the initial impulse to start inflation~\cite{inflation}. This is the case if milder potentials, which could arise from lower--dimensional branes, are taken into account~\cite{cmbred,cond_dud}. The generic indication of these scenarios, if a short inflation happened to have left us some glimpses of its inception, resonates with the lack of power displayed by the first CMB multipoles~\cite{CMB_test}. Our discussion will make these considerations a bit more concrete from a top--down perspective.

There is an extensive activity aimed at identifying and cutting out, within the vacua that arise in Supergravity, wide subsets that should not admit ultraviolet completions, and therefore should not pertain to String Theory proper~\cite{weak_gravity}. There is also a widespread feeling that instabilities show up generically in non--supersymmetric contexts. When this is the case perturbations, even if initially small, grow generically in time, and a pathology of this type is precisely what the negative squared masses of tachyons signal in flat space. In $AdS_d$ vacua matters are a bit subtler, and the proper stability conditions, which are called Breitenlohner--Freedman (BF) bounds~\cite{BF}, allow for finite ranges of negative squared masses, in ways that depend on the dimension $d$ and on the nature of the fields. Hence, one can study perturbative stability in $AdS$ backgrounds analyzing the eigenvalues of (properly defined) squared mass matrices for the available modes. This will be a primary task of this paper, and in particular we shall confirm the instability found in~\cite{gubsermitra} for the $AdS_3 \times S^7$ family of vacua, which were actually first considered there, and we shall also present a similar result for the $AdS_7 \times S^3$ heterotic vacua of~\cite{ms17}. The instabilities present in both cases comply with the general expectations in~\cite{weak_gravity}, but the relative simplicity of these settings allows us to move further.

{ In the $AdS \times S$ vacua of~\cite{gubsermitra,ms17} the dilaton has constant profiles. This makes it possible to scan the behavior of perturbations in a wide class of systems simply modifying three constants, the vacuum value $V_0$ of the potential and those of its first derivative $V_0'$ and of its second derivative $V_0''$. One can explore, in this fashion, whether perturbative instabilities are generic within the potentials that give rise to similar $AdS \times S$ flux vacua. The lesson that we shall gather is clearcut: \emph{close to the actual values, and as soon as $V_0$ becomes negative, there are wide regions of perturbative stability}. This interesting phenomenon occurs both in the $AdS_3 \times S^7$ orientifold vacua and in the $AdS_7 \times S^3$ heterotic ones, and the nearby regions could play a role when string corrections are taken into account, although we are unable to justify dynamically the emergence of such deformations via quantum corrections. Moreover, in the $AdS_7 \times S^3$ heterotic vacua the instability affects only the $\ell=1$ scalar modes, so that it can be eliminated by an antipodal projection in the internal sphere. We also explored similar, albeit more complicated, operations for the orientifold vacua. In particular, in the simpler case of an internal $S^3$ one could work with unit quaternions, removing all unwanted spherical harmonics with $\ell=2,3,4$ via projections based on the symmetry group of the cube in $\mathbb{R}^3$. This would leave no fixed sub--varieties, and therefore constitutes an alternative option for the heterotic case. Its direct counterpart in $S^7$, however, appears much more complicated, since it would rest on octonions. We thus explored, as an alternative, a construction based on three quaternion pairs, which does eliminate all unwanted spherical harmonics but unfortunately fixes some sub--varieties.}

The following sections deal with the nine--dimensional vacua with Poincar\'e symmetry and the cosmological solutions that were discovered in~\cite{dm_9Dsolution}, and also with the linear--dilaton systems of~\cite{lindil_1,lindil_2} and other generalizations. As we shall see, the nine--dimensional vacua are perturbatively stable solutions of Einstein's equations although, from the vantage point of String Theory, they include regions of high curvature and strong coupling, where corrections to Supergravity are expected to play an important role. Some features of these higher--derivative couplings were explored in~\cite{cond_dud}. In the spirit of the preceding extensions and in view of their potential applications, we also analyze the cosmological solutions that arise, in lower dimensions, with similar potentials $e^{\widetilde{\gamma}\,\phi}$, for different values of $\widetilde{\gamma}$. These potentials were first studied in~\cite{russo}, and this will also connect the present analysis to the work in~\cite{climbing}. The resulting indications are that \emph{perturbations are well-behaved, up to an intriguing behavior of the homogeneous, ${\mathbf k}=0$ mode that manifests itself, in ten dimensions, with the potentials of eqs.~\eqref{pot32} and \eqref{pot52} in the absence of milder terms, to which we shall return in Section~\ref{sec:tensor_climbing}}.

The contents of this paper are as follows. In Section \ref{sec:model} we define the low--energy Lagrangians of interest and we explain our conventions. In Section~\ref{sec:orientifold_flux} we review the $AdS_3 \times S^7$ orientifold vacua of~\cite{gubsermitra,ms17} and extend them in the most general way that is of interest with constant dilaton profiles, allowing for generic values of the potential $V_0$, its first derivative $V_0'$ and its second derivative $V_0''$. In Section~\ref{sec:perturbations37_gen} we linearize the field equations and set up the perturbative analysis, which is carried out in Section~\ref{sec:tens_vec_37} for tensor and vector perturbations. No instabilities are found in these sectors. The lowest tensor modes arise from metric perturbations, while the lowest vector modes are the expected massless Kaluza--Klein vectors for the internal $S^7$, which here arise from mixings of metric and form contributions. In Section~\ref{sec:scalar_pert_37} we analyze scalar perturbations. Here we confirm the results in~\cite{gubsermitra}: there are no instabilities in the $\ell=0$ sector, where the available modes arise from the metric tensor and the dilaton, while there are instabilities for $\ell=2,3,4$, where also the form field comes into play. However, we display a wide corner of modified potentials, which lie close to the lowest--order one in eq.~\eqref{pot32} but have a negative $V_0$, where no unstable modes are present. At the end of the section we point out how the instabilities could be removed via projections in the internal sphere for the original potential~\eqref{pot32}, although all examples that we have constructed feature fixed sub--varieties. In Section~\ref{sec:heterotic_vacuum} we review the $AdS_7\times S^3$ heterotic vacua of~\cite{ms17} and extend them in the most general way that is of interest with constant dilaton profiles, allowing again for generic values of the potential $V_0$, its first derivative $V_0'$ and its second derivative $V_0''$. In Section~\ref{sec:perturbations73_gen} we linearize the field equations and set up the perturbative analysis, which is carried out in Section~\ref{tens_vec_73} for tensor and vector perturbations. Again, no instabilities are found in these sectors. The lowest tensor and vector modes are massless spin--two excitations and the expected massless Kaluza--Klein vectors for the internal $S^7$, which here arise, again, from mixed metric and form contributions. In Section~\ref{sec:scalar_pert_73} we analyze scalar perturbations and show that there are no instabilities in the $\ell=0$ sector, where only the metric tensor and the dilaton enter, while there is again an instability when the form field comes into play. Here it only affects the $\ell=1$ modes, and could be removed by an antipodal $\mathbb{Z}_2$ orbifold projection in the internal manifold, along with all odd--$\ell$ harmonics. As for the orientifold vacua, we display however a wide corner of modified potentials that lie close to the lowest--order one in eq.~\eqref{pot32} and have a negative $V_0$, where again no unstable modes are present. In Section~\ref{sec:dm_vacua} we review the vacua with nine--dimensional Poincar\'e symmetry of~\cite{dm_9Dsolution}, and in Section~\ref{sec:pert_9D_vacuum} we set up the perturbative analysis. Then in Section~\ref{sec:scalar9d} we show that there are no perturbative instabilities in the scalar sector for the potentials in eqs.~\eqref{pot32} and \eqref{pot52}, and in Section~\ref{tensor_9d} we show that the same is true for vector and tensor perturbations.
This background is therefore a perturbatively stable solution of Einstein's equations, although from the vantage point of String Theory it includes strong--coupling regions. In Section~\ref{sec:subcritical} we analyze the stability of the linear--dilaton vacua that originate, below the critical dimension of String Theory, from~\cite{polyakov,lindil_1}.
In Section~\ref{sec:climbing} we review the salient features of the climbing scalar cosmologies that arise with the potentials of eqs.~\eqref{pot32} and \eqref{pot52}, and in Sections~\ref{sec:tensor_climbing} and \ref{sec:scalar_climbing} we analyze the corresponding perturbations. {We show that there is a logarithmic instability for the homogeneous ${\mathbf{k}}=0$ tensor mode, but there are none for other modes}. In Section \ref{sec:general_exponential} we consider the behavior of a class of milder exponential potentials, which can arise in lower dimensions, in the presence of branes, and we show that there are no instabilities for them. Finally, in Section~\ref{sec:lin_dil} we analyze a special class of potentials related to the linear--dilaton cosmologies of~\cite{lindil_2}, while Section~\ref{sec:conclusions} contains a short summary of our results and some indications on possible further developments. There are also four Appendices. In Appendix~\ref{sec:app1} we discuss some differential equations that are encountered in Sections \ref{sec:scalar_pert_37} and \ref{sec:scalar_pert_73}, while in Appendix~\ref{sec:app2} we review some properties of tensor spherical harmonics that are used extensively in Sections \ref{sec:perturbations37_gen} and \ref{sec:perturbations73_gen}. In Appendix~\ref{sec:app3} we collect some useful results on BF bounds, along with a sketchy derivation of them, and finally in Appendix~\ref{sec:app4} we collect some observations on cosmic and conformal time that are relevant for Section \ref{sec:general_exponential}.

\section{The Models} \label{sec:model}
\vspace{10pt}

In this paper we explore the low--energy dynamics of the three ten--dimensional string models with broken Supersymmetry and no tachyons in their spectra. The first two models are the $USp(32)$ orientifold with BSB of~\cite{bsb} and the $U(32)$ orientifold of \cite{u32}, whose low--energy effective field theories are identical, insofar as the vacua addressed here are concerned, while the third is the $SO(16) \times SO(16)$ heterotic model of \cite{so16xso16}. In all cases we shall account for their exponential dilaton potentials, whose origin is however different in the orientifold and heterotic examples. In the former it reflects residual brane/orientifold tensions, and is thus an open--string (projective)disk--level effect, while in the latter it arises from the torus and is therefore a standard first quantum correction. As we have shown in~\cite{ms17}, these potentials allow $AdS_3 \times S^7$ flux vacua in the two orientifold models and $AdS_7 \times S^3$ flux vacua in the heterotic model. In both classes of vacua the dilaton acquires constant values, which can be tuned into regions of small string coupling.

We use the ``mostly plus'' signature and the following definition of the Riemann curvature in terms of the Christoffel connection $(M,N=0,..9)$:
\beq
{R^A}_{B M N}(\Gamma) \ = \ \partial_N \, \Gamma_{B M}^A \ - \ \partial_M \, \Gamma_{B N}^A \ + \ \Gamma_{N C}^A \, \Gamma_{B M}^C \ - \ \Gamma_{M C}^A \, \Gamma_{B N}^C \ .
\eeq
Moreover
\beq
R_{B N} \ = \ {R^A}_{B A N}(\Gamma)
\eeq
is the Ricci tensor. Notice that with this definition the curvature of a sphere is \emph{negative}, and that the relation to ${R_{M N}}^{A B}(\omega)$, the curvature in the frame formalism, is
\beq
{R_{M N}}^{A B}(\omega) \ = \ - \ {R^{A B}}_{M N}(\Gamma) \ .
\eeq
Here we refer to the curvature defined in terms of $\Gamma$, and in our conventions
\beq
[ \nabla_M, \nabla_N] V_Q \ = \ {R^P}_{Q M N}(\Gamma)\, V_P \ .
\eeq

The low--energy dynamics of the systems of interest is captured by a class of Lagrangians involving the metric tensor, a scalar field $\phi$ and a two--form gauge field of field strength $H_3$. In the Einstein frame, these read
\beq
{\cal S}  \ = \  \frac{1}{2\,k_{10}^2}\int d^{10}x\sqrt{-g}\left[-\ R\ - \frac{1}{2}\ (\partial\phi)^2\ - \ \frac{1}{12}\ e^{\beta\,\phi}\, {\cal H}^2  \ - \ V(\phi) \,
\right] \ , \label{lagrangian_bsb}
\eeq
where in the orientifold examples, which we shall study with reference to the $USp(32)$ model with BSB and to the $U(32)$ model, $\beta=1$ and
\beq
V \ = \ T \, e^{\,\frac{3}{2} \, \phi} \ . \label{lowest_bsb}
\eeq
On the other hand, in the heterotic $SO(16) \times SO(16)$ model $\beta=-1$ and
\beq
V \ = \ T\, e^{\,\frac{5}{2} \, \phi} \ . \label{lowest_het}
\eeq

The equations of motion read, in general,
\bea
R_{MN} \ - \ \frac{1}{2}\ g_{MN}\, R &=& - \ \frac{1}{2}\  \pr_M\phi\, \pr_N\phi\ -\ \frac{1}{4}\ e^{\,\beta\,\phi} \ \left(H^2\right)_{M N} \nonumber\\
&+& \frac{1}{2}\,g_{MN}\Big[\frac{1}{2}\,(\pr\phi)^2\ +\ \frac{e^{\,\beta\,\phi}}{12}\,H^2\ +\ V(\phi)\Big] \ ,  \nonumber
\\
\Box\phi &=& \frac{\beta}{12} \ e^{\,\beta\,\phi}\ H^2\ + \ V^\prime(\phi)  \ , \label{eqsbeta}
\\
d\Big(e^{\beta\,\phi}\ {}^{*}H\Big) &=& 0 \ . \nonumber
\eea

Equivalently, one can combine the Lagrangian Einstein equation with its trace and work with
\beq
R_{MN} \ + \ \frac{1}{2} \, \pr_M\phi\, \pr_N\phi\ + \ \frac{1}{4}\, e^{\,\beta\,\phi} \, \left(H^2\right)_{M N} \ - \ g_{MN}\Big[ \frac{1}{48}\ e^{\,\beta\,\phi} \,H^2\ - \ \frac{1}{8}\ V(\phi)\Big] \ = \ 0 \ . \label{non_lag_Einstein}
\eeq
The orientifold $AdS_3 \times S^7$ vacua of interest are supported by a flux of the three--form field strength $H_{\mu\nu\rho}$ in the $AdS$ factor, while their heterotic counterparts are supported by a flux of the three--form field strength $H_{ijk}$ in the $S^3$ factor.

Actually, it proved very instructive to allow generic values, in the vacuum, of the potential $V_0$ and of its first and second derivatives $V_0'$ and $V_0''$. As we shall see, while instabilities are present, among the available scalar perturbations, in the uncorrected vacua corresponding to eqs.~\eqref{lowest_bsb} and \eqref{lowest_het}, nearby interesting regions of stability exist with $V_0 <0$. We also extended the discussion to more general systems, allowing generic values of $\beta$, but this did not add significant novelties to the picture.

{ The ensuing sections are devoted to the study of perturbations of the two classes of $AdS \times S$ backgrounds described in the Introduction. In the following, we shall thus refer all covariant derivatives to the corresponding background metrics. Moreover, we shall distinguish space--time d'Alembertians, denoted as usual by $\Box$, and internal Laplace operators, here denoted by $\nabla^2$, since we shall decompose all perturbations in the proper sets of internal spherical harmonics, on which $\nabla^2$ has the eigenvalues discussed in Appendix~\ref{sec:app2}.}

\section{The (Generalized) $AdS_3 \times S^7$ Orientifold Flux Vacua} \label{sec:orientifold_flux}
The background manifold is in this case $AdS_3\times S^7$, with metric
\beq
ds^{\,2 \, (0)}\ = \ R_{AdS}^2 \, \lambda_{\mu\nu}\, dx^\mu\, dx^\nu \ + \ R^2 \, \gamma_{ij}\, dy^i\, dy^j \ .
\eeq
Here $\lambda$ and $\gamma$ are $AdS_3$ and $S^7$ metrics of unit radius, while $R_{AdS}$ and $R$ denote the actual $AdS_3$ and $S^7$ radii. Greek indices run from 0 to 2, while Latin indices run from 1 to 7, the dilaton has a constant vacuum value $\phi_0$, and finally there is  three--form flux in $AdS$, with
\beq
H_0^{\mu\nu\rho} \ = \ \epsilon^{\mu\nu\rho} \ \frac{ {\widetilde h}}{\sqrt{-\lambda}} \ . \label{37formflux}
\eeq
Here ${\widetilde h}$ is a constant and $\epsilon^{012}\,=\,1\,=\,-\,\epsilon_{012}$, while all other components of $H^{MNP}$ vanish in the vacuum.

For maximally symmetric spaces, in terms of $R(\Gamma)$,
\bea
R^{(0)}{}_{\mu\nu\rho\sigma} &=& \ \frac{1}{R_{AdS}^2}\Big[\lambda_{\mu\rho}\, \lambda_{\nu\sigma} \ -\ \lambda_{\mu\sigma}\, \lambda_{\nu\rho}\Big]\ , \nonumber \\
R^{(0)}{}_{ijkl} &=&- \ \frac{1}{R^2}\Big[\gamma_{ik}\, \gamma_{jl} \ -\ \gamma_{jk}\, \gamma_{il}\Big]
\ ,
\eea
where $R$ and $R_{AdS}$ are the radii of the sphere and $AdS$ factors. Moreover, the preceding conventions imply that
\bea
\big[ \nabla_\mu \,,\, \nabla_\nu\big] \, V_\rho &=& \frac{1}{R_{AdS}^2} \, \Big( \lambda_{\nu\rho}\, V_\mu \ - \ \lambda_{\mu\rho}\, V_\nu \Big) \ , \nonumber \\
\big[ \nabla_i \,,\, \nabla_j\big] \, V_k &=& - \ \frac{1}{R^2} \, \Big( \gamma_{j k}\, V_i \ - \ \gamma_{i k}\, V_j \Big) \ ,
\eea
where, here and in the following Section~\ref{sec:scalar_pert_37}, covariant derivatives are computed referring the $AdS_3 \times S^7$ background. In addition, the zeroth--order dilaton equation gives
\bea
V_0^\prime \ =\ \ \frac{\beta}{2}\ {\widetilde h}^{\,2}\, e^{\,\beta\,\phi_0} \ ,
\eea
which links the three--form flux, sized by ${\widetilde h}$, to the derivative of the scalar potential. Notice that the allowed signs of $V_0'$ and $\beta$ must coincide, a condition that holds for the perturbative orientifold vacuum, where $\beta=1$.
The Einstein equations translate into
\bea
\frac{21}{R^2} \ - \ \frac{1}{R_{AdS}^2} & = & \frac{1}{4} \, e^{\,\beta\,\phi_0} \, {\widetilde h}^2 \ + \ \frac{1}{2} \ V_0 \ , \\
\frac{15}{R^2} \ - \ \frac{3}{R_{AdS}^2} & = & - \ \frac{1}{4} \, e^{\,\beta\,\phi_0} \, {\widetilde h}^2 \ + \ \frac{1}{2} \ V_0 \ ,
\eea
and it is convenient to define the two variables
\beq
\sigma_3 \ = \frac{R_{AdS}^2}{2\,\beta}\, V_0^{'} \ =  \ 1 \ + \ 3\, \frac{R_{AdS}^2}{R^2} \ , \qquad \tau_3 \ = \ R_{AdS}^2 \, V_0^{''} \ , \label{sigma3tau3}
\eeq
which will often appear in the next section. Notice that $\sigma_3 \geq 1$ and
\beq
R_{AdS}^2\,V_0 \ = \ 12 \left( \sigma_3 \ - \ \frac{4}{3} \right) \ ,
\eeq
so that the value $\sigma_3 = \frac{4}{3}$ separates negative and positive values of $V_0$ for these generalized $AdS_3 \times S^7$ vacua, and for the \emph{(projective)disk--level} orientifold potential
\beq
\sigma_3 \ = \ \frac{3}{2} \ , \qquad \tau_3 \ = \ \frac{9}{2} \ .
\eeq
\subsection{Perturbations of the Generalized $AdS_3 \times S^7$ Orientifold Flux Vacua} \label{sec:perturbations37_gen}

We can now discuss perturbations in these vacua, letting
\beq
g_{MN} = g^{(0)}_{MN} + h_{MN} \ , \quad \phi = \phi_0 + \varphi \ , \quad B_{MN}=B_{0\,MN}\ +\ e^{\,-\,\beta\,\phi_0} \, \frac{b_{MN}}{{\widetilde h}} \ ,
\eeq
and linearizing the resulting equations of motion. The perturbed tensor equations are
\bea
&\Box_{10}& \!\!\! b_{\mu\nu} \ - \ \nabla_\mu \Big( \nabla^M \, b_{M\nu} \Big) \ - \ \nabla_\nu \Big( \nabla^M \, b_{\mu M} \Big) \ + \ \frac{2}{R_{AdS}^2} \ b_{\mu\nu} \nonumber \\
&+& 4 \left( \frac{1}{R_{AdS}^2} \ + \ \frac{3}{R^2} \right)\ \sqrt{-\lambda}\ \epsilon_{\mu\nu\rho}\, \Big( \beta\,\nabla^\rho\,\varphi \ - \ \nabla^i \, {h_i}^\rho \ - \ \frac{1}{2} \ \nabla^\rho\, \lambda \cdot h \ + \ \frac{1}{2} \ \nabla^\rho\, \gamma \cdot h \Big) \ = \ 0 \ , \nonumber \\
&\Box_{10}& \!\!\!  b_{\mu i} \ - \ \nabla_\mu \Big( \nabla^M \, b_{M i} \Big) \ - \ \nabla_i \Big( \nabla^M \, b_{\mu M} \Big) \ + \ \Big( \frac{2}{R_{AdS}^2} \ - \ \frac{6}{R^2} \Big)\,  b_{\mu i}  \\
&+& 4 \left( \frac{1}{R_{AdS}^2} \ + \ \frac{3}{R^2} \right) \ \sqrt{-\lambda}\ {\epsilon_{\alpha\beta\mu}}\, \nabla^\alpha\, {h^{\beta}}_i \ = \ 0 \ , \nonumber \\
&\Box_{10}& \!\!\!  b_{i j} \ - \ \nabla_i \Big( \nabla^M \, b_{Mj} \Big) \ - \ \nabla_j \Big( \nabla^M \, b_{iM} \Big) \ - \ \frac{10}{R^2} \ b_{ij} \ = \ 0 \ , \nonumber
\eea
where, here and in the following,
\beq
\Box_{10} \ = \ \Box \ + \ \nabla^2 \ , \label{box_decompose}
\eeq
in terms of the $AdS$ and sphere contributions.
In a similar fashion, the perturbed dilaton equation is
\beq
\Box_{10}\, \varphi \ - \ V_0^{\prime\prime}\, \varphi \ + \ 2 \left( \frac{1}{R_{AdS}^2} \ + \ \frac{3}{R^2} \right) \left(\beta^2\, \varphi   \ - \ \beta\,\lambda \cdot h \right)
\ - \ \frac{\beta}{2}\ \frac{\epsilon^{\,\mu\nu\rho}}{\sqrt{-\lambda}} \, \nabla_{\mu}\,b_{\nu\rho}\ = \ 0 \ .
\eeq
Finally, the perturbed metric equations that follow from
eq.~\eqref{non_lag_Einstein} rest on the linearized Ricci tensor
\bea
R_{NR}&=& R^{(0)}_{NR}\ +\ \frac{1}{2}\, \Big[\Box\, h_{NR} \ -\ \nabla_N \left(\nabla \cdot h\right)_{R}\ - \ \nabla_R \left(\nabla \cdot h\right)_{N}\ +\ \nabla_N\nabla_R \, h_L{}^L \Big] \nonumber \\
&+& \frac{1}{2} \ {R^{\,(0)A}}_N \ h_{AR} \ + \ \frac{1}{2} \ {R^{\,(0) A}}_R \ h_{AN} \ -\ {{{R^{\,(0) A}}_N}^B}_R\ h_{AB} \ , \label{pert_RMN}
\eea
and read
\bea
\Box_{10} \, h_{\mu\nu} &+& \frac{2}{R_{AdS}^2} \, h_{\mu\nu} \ -\ \nabla_\mu \left(\nabla \cdot h\right)_{\nu}\ - \ \nabla_\nu \left(\nabla \cdot h\right)_{\mu}\ +\ \nabla_\mu\,\nabla_\nu \, (\lambda \cdot h + \gamma \cdot h) \nonumber \\ &+& \lambda_{\mu\nu}\Big[ -\, \frac{5\,\beta}{2} \left(\frac{1}{R_{AdS}^2} + \frac{3}{R^2} \right) \varphi  + 3\,\lambda \cdot h \left(-\,\frac{1}{R_{AdS}^2} + \frac{3}{R^2} \right) \, - \, \frac{3}{4}\, \frac{\epsilon^{\alpha\beta\gamma}}{\sqrt{-\lambda}} \, \nabla_{\alpha}\,b_{\beta\gamma}\Big]  \ = \ 0 \ ,\nonumber \\
\Box_{10} \, h_{\mu i} &+&  h_{\mu i} \left( \frac{2}{R_{AdS}^2} + \frac{6}{R^2} \right) \ -\ \nabla_\mu \left(\nabla \cdot h\right)_{i}\ - \ \nabla_i \left(\nabla \cdot h\right)_{\mu}\ +\ \nabla_\mu\,\nabla_i \, (\lambda \cdot h + \gamma \cdot h)\nonumber \\ &+&  \,\frac{1}{2\,\sqrt{-\lambda}} \ {\epsilon}^{\alpha\beta\gamma} \, \lambda_{\mu\gamma} \, (\nabla_i\,b_{\alpha\beta}+\nabla_\alpha\,b_{\beta i}+\nabla_\beta\,b_{i \alpha} )  \ = \ 0 \ , \label{tensor_bsb}\\
\Box_{10} \, h_{i j} &-& \frac{2}{R^2}\ h_{ij} \ -\ \nabla_i \left(\nabla \cdot h\right)_{j}\ - \ \nabla_j \left(\nabla \cdot h\right)_{i}\ +\ \nabla_i\,\nabla_j \, (\lambda \cdot h + \gamma \cdot h) \nonumber \\ &+&  \gamma_{i j}\left[  \frac{2}{R^2} \ \gamma \cdot h \ + \
\left( \frac{1}{R_{AdS}^2} \,+\, \frac{3}{R^2}\right) \left(\frac{3}{2}\,\beta\,\varphi \,- \,\lambda \cdot h\right) \ - \ \frac{\epsilon^{\alpha\beta\gamma}\,\nabla_{\alpha}\,b_{\beta\gamma}}{4\,\sqrt{-\lambda}}  \right] \ = \ 0 \ . \nonumber
\eea

In all cases, perturbations depend on the $AdS$ coordinates $x^\mu$ and on the sphere coordinates $y^i$, and will be expanded in corresponding spherical harmonics, whose structure is briefly reviewed in Appendix~\ref{sec:app2}. For instance, for internal scalars this will always result in expressions of the type
\beq
h_{\mu\nu}(x,y) \ = \ \sum_\ell\, h_{\mu\nu,\,I_1 \ldots I_\ell}(x) \, {\cal Y}_{(8)}^{I_1\dots I_\ell}(y)\ ,
\eeq
where $I_i=1,\ldots ,8$ and $h_{\mu\nu,\,I_1 \ldots I_\ell}(x)$ is totally symmetric and traceless in the Euclidean $I_i$ labels. However, the eigenvalues of the internal Laplace operator $\nabla^2$ will only depend on $\ell$. Hence, for the sake of brevity, and at the cost of being somewhat sketchy, we shall leave the internal labels implicit, although in some cases we shall refer to their ranges when counting multiplicities.

For tensors in internal space there are some additional complications. For example, for mixed metric components the expansion reads
\beq
h_{\mu i}(x,y) \ = \ \sum_\ell\, h_{\mu J,\,I_1 \ldots I_\ell}(x) \ {\cal Y}_{(8)\, i}^{I_1\dots I_\ell;J}(y)
\eeq
where $h_{\mu J,\,I_1 \ldots I_\ell}(x)$ corresponds to a ``hooked'' Young tableau of mixed symmetry and $\ell \geq 1$, as explained in Appendix B. Here the ${\cal Y}_{(8)\,i}$ are vector spherical harmonics, and we shall drop all internal labels, for brevity, also for the internal tensors that we shall consider.

\subsection{Tensor and Vector Perturbations} \label{sec:tens_vec_37}

Following standard practice, we classify perturbations referring to their behavior under the isometry group $SO(2,2) \times SO(8)$ of the $AdS_3 \times S^7$ background. In this fashion, the possible unstable modes violate the $BF$ bounds, which depend on the nature of the fields involved and correspond, in general, to finite negative values for (properly defined) squared $AdS$ masses. Indeed, as reviewed in Appendix~\ref{sec:app3} with reference to forms, care must be exercised to identify the proper masses to which the bound applies, since in general they differ from the eigenvalues of the corresponding $AdS$ d'Alembertian. In particular, aside from the case of scalars, massless field equations always exhibit gauge symmetries.

Let us begin by considering tensor perturbations, which result from transverse traceless $h_{\mu\nu}$, with all other perturbations vanishing.
The corresponding dynamical equation
\beq
\Box \, h_{\mu\nu} \ - \ \frac{\ell(\ell+6)(\sigma_3-1)}{3\, R_{AdS}^2}\ h_{\mu\nu} \ +\  \frac{2}{R_{AdS}^2} \, h_{\mu\nu} \ = \ 0 \ \qquad ( \ell \geq 0) \ , \label{gravity_37}
\eeq
where we have replaced the internal radius $R$ with the $AdS$ radius $R_{AdS}$ using eq.~\eqref{sigma3tau3}, obtains when the first of \eqref{tensor_bsb} is combined with the results summarized in Appendix \ref{sec:app2} on general spherical harmonics. These harmonics are eigenvectors of the internal Laplace operator in eq.~\eqref{box_decompose}, whose eigenvalues on $SO(8)$ scalars are $- \frac{\ell(\ell+6)}{R^2}$, with $\ell=0,1,...$ an integer number.
In order to properly interpret this result, however, it is crucial to notice that the massless tensor equation in $AdS$ is the one determined by gauge symmetry. In fact, the linearized Ricci tensor determined by eq.~\eqref{pert_RMN} is \emph{not} gauge invariant under linearized diffeomorphisms of the $AdS$ background, but
\beq
\delta\,R_{\mu\nu} \ = \ \frac{2}{R_{AdS}^2} \left(\nabla_\mu\,\xi_\nu \ + \ \nabla_\nu\,\xi_\mu \right) \ .
\eeq
However, the fluxes present in eq.~\eqref{non_lag_Einstein} endow, consistently, its \emph{r.h.s.} with a similar behavior, and $\ell=0$ in eq.~\eqref{gravity_37} corresponds precisely to massless modes. Thus, as expected from Kaluza--Klein theory, eq.~\eqref{gravity_37} describes a massless field for $\ell=0$, and an infinity of massive ones for $\ell > 0$. These perturbations are all consistent with the $BF$ bound, and therefore \emph{no instabilities} are present in this sector.

There are also scalar excitations resulting from the traceless part of $h_{ij}$ that is also divergence--free, which are tensors with respect to the internal rotation group. They satisfy (see Appendix~\ref{sec:app2})
\beq
\Box \, h_{i j} \ - \ \frac{\ell(\ell+6)(\sigma_3-1)}{3\,R_{AdS}^2}\ h_{ij} \ = \ 0 \qquad ( \ell \geq 2)\ ,
\eeq
so that their squared masses are all positive. Finally, there are massive $b_{ij}$ perturbations, which are divergence--free and satisfy
\beq
\Box\,b_{i j} \ - \ \frac{[\ell(\ell+6)+8](\sigma_3-1)}{3\,R_{AdS}^2} \ b_{ij} \ = \ 0 \qquad ( \ell \geq 2)\ .
\eeq

{ Vector perturbations are a bit more involved, due to mixings between $h_{\mu i}$ and $b_{\mu i}$ induced by fluxes. The relevant equations are
\bea
\Box_{10}\,  b_{\mu i} &+& \Big( \frac{2}{R_{AdS}^2} \ - \ \frac{6}{R^2} \Big)\,  b_{\mu i}  \ +\ 4 \left( \frac{1}{R_{AdS}^2} \ + \ \frac{3}{R^2} \right) \ \sqrt{-\lambda}\ {\epsilon_{\alpha\beta\mu}}\, \nabla^\alpha\, {h^{\beta}}_i \ = \ 0 \ , \nonumber \\
\Box_{10} \, h_{\mu i} &+&  \Big( \frac{2}{R_{AdS}^2} + \frac{6}{R^2} \Big)h_{\mu i} \ +\ \frac{1}{2\,\sqrt{-\lambda}} \ {\epsilon}^{\alpha\beta\gamma} \, \lambda_{\mu\gamma} \, (\nabla_\alpha\,b_{\beta i}+\nabla_\beta\,b_{i \alpha} )  \ = \ 0 \ , \label{vect_pert}
\eea
where $h_{\mu i}$ and $b_{\mu i}$ are divergence--free in both indices. It is now possible to write
\beq
b_{\mu i} \ = \ \sqrt{-\lambda}\ {\epsilon_{\alpha\beta\mu}}\, \nabla^\alpha\, {F^{\beta}}_i \ , \label{bmui}
\eeq
but this does not determine $F_i^\beta$ uniquely, since the redefinitions
\beq
{F^{\beta}}_i  \ \rightarrow \ {F^{\beta}}_i  \ + \ {\nabla^\beta}\, \Lambda_i  \label{Fgauge}
\eeq
do not affect $b_{\mu i}$. The divergence--free $b_{\mu i}$ of interest obtain, in particular, provided $F_i^\beta$ is divergence--free in its internal index $i$, and divergence--free $\Lambda_i$ do not affect this condition.

One is thus led to the system~\footnote{In all these expressions that refer to vector perturbations $\ell \geq 1$, as described in Appendix~\ref{sec:app2}.}
\bea
\Box\,  {F_i}^\mu &+&  \Big[ \frac{2}{R_{AdS}^2} \ - \ \frac{\ell(\ell+6)+5}{R^2} \Big]\,  {F_i}^\mu  \ + \ 4 \left( \frac{1}{R_{AdS}^2} \ + \ \frac{3}{R^2} \right) {h_i}^\mu \ = \ 0 \ , \nonumber \\
\Box\,  {h_i}^\mu &+& \frac{\ell(\ell+6)+5}{R^2} \ {F_i}^\mu \ - \ \Big[ \frac{2}{R_{AdS}^2} \ + \ \frac{\ell(\ell+6)+5}{R^2} \Big]\,  {h_i}^\mu  \ = \ 0 \ \label{bheqs}
 .
\eea
and the first of eqs.~\eqref{bheqs} could in principle accommodate a source term of the type $\nabla^\mu\, {\widetilde \Lambda}_i$. However, its contribution can be absorbed by a gauge redefinition of the type~ \eqref{Fgauge}, and from now on we shall ignore it. Similar arguments apply to the ensuing analysis of scalar perturbations in this section and to some cases in Section~\ref{sec:heterotic_vacuum}.}

In terms of the combination $\sigma_3$ of eq.~\eqref{sigma3tau3} this system becomes
\bea
R_{AdS}^2\,\Box\,  {F_i}^\mu &+&  \Big[ 2 \ - \ \frac{\ell(\ell+6)+5}{3}\ (\sigma_3 - 1) \Big]\,  {F_i}^\mu  \ {+} \ 4 \, \sigma_3\, {h_i}^\mu \ = \ 0 \ , \\
R_{AdS}^2\,\Box\,  {h_i}^\mu &+& \frac{\ell(\ell+6)+5}{3} \ (\sigma_3-1) \ {F_i}^\mu \ - \ \Big[2 \ + \ \frac{\ell(\ell+6)+5}{3}\, (\sigma_3-1)\Big]\,{h_i}^\mu   \ = \ 0 \
 . \nonumber
\eea

The eigenvalues of the mass matrix are thus
\beq
(\sigma_3-1)\ \frac{\ell(\ell+6)+5}{3} \, \pm  {2}\ \sqrt{1 \ +  \ \sigma_3(\sigma_3-1)\ \frac{\ell(\ell+6)+5}{3}} \ .
\eeq
In order to refer to the BF bound in Appendix~\ref{sec:app3}, one should add 2 to these expressions and compare the result with zero. All in all, \emph{there are no modes below the BF bound in this sector.} The vector modes lie above it for $\ell>1$ in the region $\sigma_3>1$, while they are massless for $\ell=1$ and all allowed values of $\sigma_3 > 1$, and also, for all $\ell$, in the singular limit $\sigma_3=1$, which would translate into a seven--sphere of infinite radius. For $\ell=1$ there are 28 massless vectors arising from one of the eigenvalues above. Indeed, according to Appendix \ref{sec:app2} they build up a second--rank antisymmetric tensor in internal vector indices, and therefore an adjoint multiplet of $SO(8)$ vectors. This counting is consistent with Kaluza--Klein theory and reflects the internal symmetry of $S^7$, although the massless vectors originate here from mixed contributions of the metric and the two--form field.

\subsection{Scalar Perturbations} \label{sec:scalar_pert_37}

Let us now focus on scalar perturbations of the complete system.
To begin with, $b_{\mu\nu}$ contributes to scalar perturbations, as can be seen letting
\beq
b_{\mu\nu} \ = \ \sqrt{- \lambda} \ \epsilon_{\mu\nu\rho}\, \nabla^\rho\, B \ , \label{ansatzB37}
\eeq
an expression that satisfies identically
\beq
\nabla^\mu\, b_{\mu\nu} \ = \ 0 \ . \label{transverse_b}
\eeq
On the other hand, they do not arise from $b_{\mu i}$ and $b_{ij}$, since the corresponding contributions would be pure gauge. On the other hand, scalar metric perturbations can be parametrized as
\bea
h_{\mu\nu} &=& \lambda_{\mu\nu}\, A \ , \nonumber \\
h_{\mu i} &=& R^2\, \nabla_\mu \, \nabla_i \, D \ , \\
h_{i j} &=& \gamma_{i j}\, C \ ,\nonumber
\eea
up to a gauge transformation with independent parameters along $AdS_3$ and $S^7$ directions. The linearized $b_{\mu\n}$ {tensor equation} yields
\beq
\Box_{10}\,B  \ + \ 4 \left( \frac{1}{R_{AdS}^2} \ + \ \frac{3}{R^2} \right) \left( \beta\,\varphi \ - \ R^2\,\nabla^2\, D \ - \ \frac{3}{2} \ A \ + \ \frac{7}{2} \ C \right) \ = \ 0 \ , \label{tensor_B_BSB}
\eeq
where we use the decomposition of eq.~\eqref{box_decompose}, so that $\nabla^2$ denotes the internal background Laplacian. After expanding in internal spherical harmonics, so that $\nabla^2 \to - \frac{\ell(\ell+6)}{R^2}$, eq.~\eqref{tensor_B_BSB} becomes finally (an $AdS$ derivative of)
\beq
\Box\,B  \ - \ \frac{\ell(\ell+6)}{R^2}\ B \ + \ 4 \left( \frac{1}{R_{AdS}^2} \ + \ \frac{3}{R^2} \right) \left( \beta\,\varphi \ + \ {\ell(\ell+6)}\,  D \ - \ \frac{3}{2} \ A \ + \ \frac{7}{2} \ C \right) \ = \ 0 \ . \label{tensor_B_final_BSB}
\eeq
{ Notice that a redefinition $B \to B+f(y)$, where $f(y)$ depends only on internal coordinates, would not affect $b_{\mu\nu}$ in eq.~\eqref{ansatzB37}. As a result, while eqs.~\eqref{tensor_B_BSB} and \eqref{tensor_B_final_BSB} could in principle contain a source term, this can be eliminated taking this fact into account. Similar considerations will apply for the $B$ field in Section~\ref{sec:scalar_pert_73}. }

In a similar fashion, the {dilaton equation} becomes
\beq
\Box\,\varphi  \ - \ \frac{\ell(\ell+6)}{R^2}\ \varphi \ - \ V_0^{\prime\prime}\, \varphi \ + \ 2 \left( \frac{1}{R_{AdS}^2} \ + \ \frac{3}{R^2} \right) \left( \beta^2\,\varphi   \ - \ 3\,\beta\,A \right)
\ + \ \beta\,\Box\, B \ = \ 0 \ ,
\eeq
or, after combining it with eq.~\eqref{tensor_B_final_BSB},
\bea
\Box\,\varphi  &-& \frac{\ell(\ell+6)}{R^2}\ \varphi \ - \ V_0^{\prime\prime}\, \varphi \ + \ \beta\,
\frac{\ell(\ell+6)}{R^2}\ B \nonumber \\ &-& \left(
\frac{1}{R_{AdS}^2} \ + \ \frac{3}{R^2} \right) \left( 2\, \beta^2\,\varphi  \,+\,
{4\,\ell(\ell+6)}\, \beta\,D \,+\,14\,\beta\,C\right) \ = \ 0 \ .
\eea

Finally, after eliminating $\Box\,B$ with eq.~\eqref{tensor_B_final_BSB}, the metric equations \eqref{tensor_bsb} read
\bea
&\lambda_{\mu\nu}& \left[ \Box\,A \,-\, \left( \frac{4}{R_{AdS}^2} \ + \ \frac{\ell(\ell+6)}{R^2} \right) A         \,+\, \left( \frac{1}{R_{AdS}^2} \ + \ \frac{3}{R^2} \right) \left( \frac{7}{2}\, \beta\,\varphi \ + \ 21\, C \ + \
{6\,\ell(\ell+6)}
\, D \right) \right.\nonumber \\ &-& \left. \frac{3\,\ell(\ell+6)}{2\,R^2}\, B \right] \ + \ \nabla_\mu\,\nabla_\nu \left[A\,+\,7\,C\,+\,2\,{\ell(\ell+6)}\, D
\right] \ = \ 0 \ , \nonumber \\                                                                              &\nabla_\mu& \!\!\!\!\!\!\nabla_i \left({12} \, D \ - \ B \ + \ 2\, A \ + \ 6\, C\right) \ = \ 0 \
,  \nonumber   \\
&\gamma_{ij}& \left\{
\Box\, C  - C \left( \frac{7}{R_{AdS}^2} + \frac{9}{R^2} +
\frac{\ell(\ell+6)}{R^2}\right) - \frac{\ell(\ell+6)}{2}\left[ 4 \left( \frac{1}{R_{AdS}^2} +
\frac{3}{R^2}\right) D \ - \ \frac{B}{R^2}\right] \right.   \nonumber \\ &-& \left. \frac{1}{2}\ \beta\,\varphi \left(
\frac{1}{R_{AdS}^2} \ + \ \frac{3}{R^2} \right) \right\}
\ + \ \nabla_i \nabla_j \left( 3\,A \
+ \ 5\, C \ - \ 2\,R^2\,\Box D \right)  \ .
\eea

These equations have an unfamiliar form, and for this reason in Appendix A we prove that the terms involving gradients must vanish separately. One ought nonetheless to distinguish two cases.

For $\ell=0$ nothing depends on internal coordinates, the terms involving $\nabla_\mu \nabla_i$ and $\nabla_i \nabla_j$ become empty and $D$ also disappears. According to Appendix A, in this case one is thus led to the reduced system
\bea
\Box\,B  &+& 4 \left( \frac{1}{R_{AdS}^2} \ + \ \frac{3}{R^2} \right)
\left( \beta\,\varphi \ + \ 14 \, C \right) \ = \ 0  \ ,    \nonumber \\
\Box\,\varphi  &-& V_0^{\prime\prime}\, \varphi \ - \ \left(
\frac{1}{R_{AdS}^2} \ + \ \frac{3}{R^2} \right) \left( 2\, \beta^2\,\varphi   \ + \ 14 \,\beta\,C\right) \ = \ 0 \ , \nonumber \\
\Box\, C  &-& C \left( \frac{7}{R_{AdS}^2} + \frac{9}{R^2}\right) \ - \  \frac{1}{2}\ \beta\, \varphi \left(
\frac{1}{R_{AdS}^2} \ + \ \frac{3}{R^2} \right)    \ = \ 0     \ , \label{eqsl01}
\eea
to be supplemented by the linear relation
\beq
A = - \,7\, C \ .
\eeq
Using the definitions of $\sigma_3$ and $\tau_3$, eqs.~\eqref{eqsl01} can be recast in the form
\bea
R_{AdS}^2\,\Box\, A  &-& \left(4 \,+\, 3\,\sigma_3\right) A \ +\ \frac{7\,\beta\,\sigma_3}{2}\ \varphi \ = \ 0  \ ,    \nonumber \\
R_{AdS}^2\, \Box\,\varphi  &+& 2\,\beta\,\sigma_3\,A \ - \ \left(\tau_3 \,+\,2\,\beta^2\,\sigma_3\right)\varphi  \ = \ 0 \ , \nonumber \\
R_{AdS}^2\, \Box\,B  &-& 8\, \sigma_3\,A  \ + \ 4\,\sigma_3\,\beta\,\varphi \ = \ 0 \ ,
\eea
and the last column of the resulting mass matrix vanishes, so that there is a vanishing eigenvalue whose eigenvector is proportional to $B$. This perturbation is however pure gauge, since eq.~\eqref{ansatzB37} implies that the corresponding field strength vanishes identically. Leaving it aside, one can work with the reduced mass matrix (in our $(\Box-{\cal M}^2)$ convention) determined by the other two equations,
\beq
R_{AdS}^2\, {\cal M}^2 \ = \  \left(
\begin {array}{cc}
4 \,+\,3\,\sigma_3 & -\, \frac{7\,\beta\,\sigma_3}{2} \\  -\,2\,\beta\,\sigma_3 \ \ & \ \ \tau_3\,+\,2\,\beta^2\,\sigma_3
\end {array}
\right)      \ , \label{mass37l0}
\eeq
whose eigenvalues are
\beq
{\beta}^{2}\sigma_3+\frac{3\sigma_3}{2}+\frac{\tau_3}{2}+2 \,\pm\, \frac{1}{2}\, \sqrt{4\,{\beta}^{4}{\sigma_3}^{2}+16\,\sigma_3
 \left( \sigma_3+\frac{\tau_3}{4}-1 \right) {\beta}^{2}+9\, \left( \sigma_3-\frac{\tau_3}{3}+\frac{4}{3} \right) ^{2}}
   \ .
\eeq

There are regions of instability as one varies the parameters, but for the actual orientifold potential, where $\left(\beta,\sigma_3,\tau_3\right)=\left(1,\frac{3}{2},\frac{9}{2}\right)$, the two eigenvalues,
\beq
\lambda_2 \ = \ \frac{12}{R_{AdS}^2} \ , \qquad  \lambda_3 \ = \ \frac{4}{R_{AdS}^2} \ ,
\eeq
are positive, and thus lie well above the Breitenlohner--Freedman bound. To reiterate, there are \emph{no unstable scalar modes for the BSB flux vacuum in the $\ell=0$ sector for the internal $S^7$.} In view of the ensuing discussion, let us add that the stability persists for convex potentials, with $\tau_3>0$, independently of $\sigma_3$.

For $\ell \neq 0$ the system becomes more complicated, since it now includes
the two algebraic constraints
\bea
&& A\,+\,7\,C\,+\,2\ {\ell(\ell+6)}\, D \ = \ 0 \ , \nonumber \\
&& {12}\, D \ - \ B \ + \ 2\, A \ + \ 6\, C \ = \ 0 \ , \label{algebr_bsb_all_l}
\eea
and the five dynamical equations
\bea
\Box\,A &-& \left( \frac{4}{R_{AdS}^2} \ + \ \frac{\ell(\ell+6)}{R^2} \right) A         \,+\, \left( \frac{1}{R_{AdS}^2} \ + \ \frac{3}{R^2} \right) \left( \frac{7}{2}\, \beta\,\varphi \ - \ 3\, A \right) \nonumber \\
&-& \frac{3\,\ell(\ell+6)}{2\,R^2}\, B \ = \ 0 \ , \nonumber \\
\Box\, C  &-& C \left( \frac{7}{R_{AdS}^2} + \frac{9}{R^2} +
\frac{\ell(\ell+6)}{R^2}\right) - \frac{\ell(\ell+6)}{2}\left[ 4 \left( \frac{1}{R_{AdS}^2} +
\frac{3}{R^2}\right) D \ - \ \frac{B}{R^2}\right]  \nonumber \\ &-& \frac{\beta}{2}\ \varphi \left(
\frac{1}{R_{AdS}^2} \ + \ \frac{3}{R^2} \right) \ = \ 0 \ , \nonumber \\
\Box\, D &-& \frac{3}{2\, R^2}\ A \ - \ \frac{5}{2\, R^2} \ C \ = \ 0 \ , \label{complete_all_l} \\
\Box\,B  &-& \frac{\ell(\ell+6)}{R^2}\ B \ + \ 4 \left( \frac{1}{R_{AdS}^2} \ + \ \frac{3}{R^2} \right) \left( \beta\,\varphi \ - \ {2} \, A \right) \ = \ 0 \ , \nonumber \\
\Box\,\varphi  &-& \frac{\ell(\ell+6)}{R^2}\ \varphi \ - \ V_0^{\prime\prime}\, \varphi \ + \
\beta\,\frac{\ell(\ell+6)}{R^2}\ B \nonumber \\ &-& \left(
\frac{1}{R_{AdS}^2} \ + \ \frac{3}{R^2} \right) \left( 2\, \beta^2\,\varphi  \,-\,
2\,\beta\,A\right) \ = \ 0 \ . \nonumber
\eea
\begin{figure}[ht]
\begin{center}
\includegraphics[width=90mm]{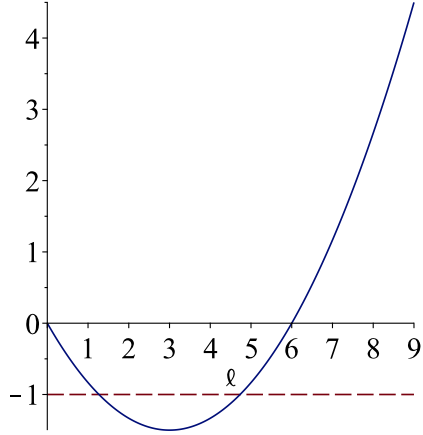}
\vspace*{-5.5truecm}
\end{center}
\caption{\small Violation of the BF bound in the BSB flux vacuum for $\beta=1$. The dangerous eigenvalue is displayed in units of $\frac{1}{R_{AdS}^2}$, and the BF bound is -1 in this case. Notice the peculiar behavior, already spotted in~\cite{gubsermitra}, whereby the squared masses \emph{decrease} initially, rather than increasing, as $\ell$ increases between 1 and 3.}
\label{fig:bad_eigenvalue}
\end{figure}

Let us first notice that this set of \emph{seven} equations for the \emph{five} unknowns $(A,B,C,D,\varphi)$ is consistent: one can indeed verify that eqs.~\eqref{complete_all_l} respect identically the algebraic constraints of eqs.~\eqref{algebr_bsb_all_l}.
One can thus concentrate on the three equations for $(A, \varphi, B)$, which do not involve other fields and read
\bea
&& R_{AdS}^{2}{\Box}\,A \ - \ \left[ 4 \ + \ 3\, \sigma_3 \ + \ \frac{L_3}{3}\ (\sigma_3 - 1) \right]A \ + \ \frac{7}{2} \ \beta\,\sigma_3 \, \varphi \ - \ \frac{L_3}{2} \ (\sigma_3 - 1) \, B \ = \ 0 \ , \nonumber \\
&& R_{AdS}^{2}{\Box}\,\varphi \ + \ 2\, \beta\,\sigma_3\, A \ - \ \left[2\, \beta^2\,\sigma_3 \ + \tau_3 \ + \ \frac{L_3}{3}\ (\sigma_3 - 1) \right]\varphi \ + \ \beta\, \frac{L_3}{3} \ (\sigma_3 - 1) \, B \ = \ 0 \ , \nonumber \\
&& R_{AdS}^{2}{\Box}\,B \ - \ 8\, \sigma_3\, A \ + \ 4 \, \beta\,\sigma_3 \ \varphi \ - \ \frac{L_3}{3} \ (\sigma_3 - 1) \, B \ = \ 0 \ ,
\eea
where we have introduced the shorthand notation
\beq
L_3 \ = \ \ell(\ell+6) \ ,
\eeq
to then determine $C$ and $D$ via the algebraic constraints.
\begin{figure}[ht]
\begin{center}
\includegraphics[width=110mm]{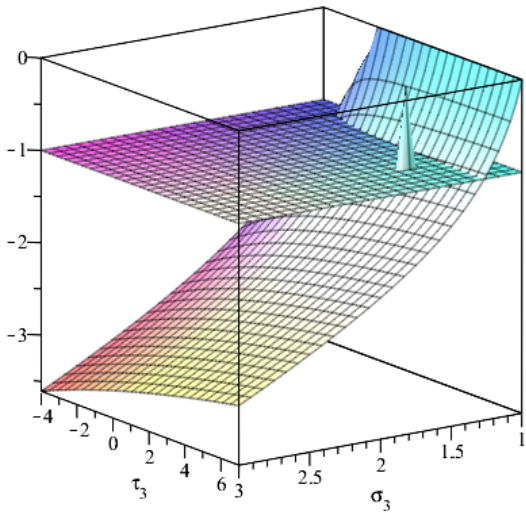}
\vspace*{-6truecm}
\end{center}
\caption{\small Comparison between the lowest eigenvalue of $R_{AdS}^2\,{\cal M}^2$ and the BF bound, which is -1 in this case. There are regions of stability for values of $\sigma_3$ close to 1, which correspond to $\frac{R^2}{R_{AdS}^2} > 9$ and negative values of $V_0$. The example displayed here refers to $\ell=3$, which corresponds to the minimum in the orientifold case of fig.~\ref{fig:bad_eigenvalue}, and the peak identifies the point corresponding to the tree--level values $\sigma_3=\frac{3}{2}$, $\tau_3=\frac{9}{2}$.}
\label{fig:bad_eigenvalue_gen}
\end{figure}

The mass matrix of interest is therefore
\beq
 R_{AdS}^2\,{\cal M}^2 \, = \, \frac{L_3}{3} \, (\sigma_3 - 1) \ {\mathbf 1}_3 \,+\,  \left( \begin {array}{ccc} 4 \, + \, 3 \sigma_3 &- \ \frac{7}{2} \ \beta\,\sigma_3 &\frac{L_3}{2} \ (\sigma_3 - 1)
\\ \noalign{\medskip} - \, 2\, \beta\,\sigma_3 & 2\, \beta^2\,\sigma_3 \, + \,\tau_3 & - \, \frac{\beta\,L_3}{3} \ (\sigma_3 - 1)
\\ \noalign{\medskip} 8\, \sigma_3 & - \, 4 \,\beta\, \sigma_3 & 0
\end {array} \right) \, , \label{m2_r}
\eeq
where now $L_3 \neq 0$. In all cases one is to compare the squared mass eigenvalues with the Breitenlohner--Freedman (BF) bound for scalar perturbations, which in this $AdS_3 \times S^7$ case reads
\beq
{\cal M}^2 \ \geq \ - \ \frac{1}{R_{AdS}^2} \ .
\eeq

For the BSB case, one is thus led, in agreement with~\cite{gubsermitra}, to the simple results
\beq
\frac{1}{R_{AdS}^2}  \left[ \begin {array}{c} \frac{\ell(\ell+6)}{6}+4\\ \noalign{\smallskip} \frac{(\ell+6)(\ell+12)}{6}\\ \noalign{\smallskip}\frac{\ell(\ell-6)}{6}\end {array} \right] \ ,
\eeq
and \emph{the BF bound is thus violated by the third eigenvalue for $\ell=2,3,4$}. Decreasing the value of $\beta$ could remove the problem for $\ell=4$, but the instability would still be present for $\ell=2,3$. On the other hand, increasing the value of $\beta$ instabilities show up also for higher values of $\ell$.
\begin{figure}[ht]
\begin{center}
\includegraphics[width=110mm]{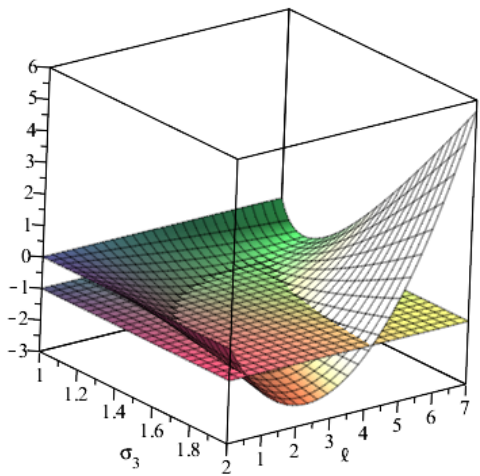}
\vspace*{-6truecm}
\end{center}
\caption{\small A different view. Comparison between the lowest eigenvalue of $R_{AdS}^2\,{\cal M}^2$ and the BF bound, which is -1 in this case, as a function of $\ell$ and $\sigma_3$, for $\tau_3=\frac{9}{2}$. There are regions of stability for values of $\sigma_3$ close to 1, which correspond to large values for the ratio $\frac{R^2}{R_{AdS}^2}$ and to negative values of $V_0$.}
\label{fig:bad_eigenvalue_different}
\end{figure}

The issue of interest is now whether there exist regions within the parameter space spanned by $\sigma_3$ and $\tau_3$ where the violation does not occur. We did find them, for all dangerous values of $\ell$, \emph{for values of $\sigma_3$ that are close to one, and therefore for negative $V_0$, and for positive $\tau_3$, {\it i.e.} for potentials that are convex close to the vacuum configuration. } These results are displayed in figs.~\ref{fig:bad_eigenvalue_gen} and \ref{fig:bad_eigenvalue_different}.

One can try to eliminate the unstable modes present for $\ell=2,3,4$ by projections in the internal $S^7$, which can be embedded in $\mathbb{C}^4$ constraining its four complex coordinates $Z^i$ to satisfy
\beq
\sum_{i=1}^4\,Z^i\,\overline{Z}_i \ = \ R^2 \ .
\eeq
According to Appendix~\ref{sec:app2}, scalar spherical harmonics of order $\ell$ are harmonic polynomials of degree $\ell$ in the $Z^i$ and their complex conjugates, so that the issue is how to project out the dangerous ones. The three--sphere, which can be embedded in $\mathbb{C}^2$ demanding that
\beq
\sum_{i=1}^2\,Z^i\,\overline{Z}_i \ = \ R^2 \ ,
\eeq
provides an instructive simpler case. One can indeed associate each point on $S^3$ to a quaternion of unit norm
\beq
Q \ = \ \left( \begin{array}{cc} Z_1 & i\,Z_2 \\ i\,\overline{Z}_2 & \overline{Z}_1  \end{array}\right)
\eeq
on which the $SU(2)$ rotations
\beq
R_i = e^{\,i\,\frac{\pi}{4}\,\sigma_i}\ , \quad R_i^8 \ = \ 1 \qquad (i=1,2,3) \label{su2_rot}
 \eeq
act freely. One can verify that these rotations, when composed in all possible ways, build the symmetry group of the cube in the three--dimensional Euclidean space associated to the three generators $\frac{\sigma_i}{2}$. One can also show that these operations suffice to eliminate all harmonic polynomials of degrees $\ell \leq 4$, while leaving no fixed sub--varieties on account of their free action on quaternions by left multiplication. One would naturally expect octonions of unit norm to play a similar role for $S^7$, but have just taken a cursory look at this construction. Alternatively, and more simply, one could consider the transformations generated by the $R_i$ in eq.~\eqref{su2_rot}  acting simultaneously on complementary pairs of $(Z^i,Z^j)$ coordinates. This would suffice to eliminate all unwanted spherical harmonics, but unfortunately it would also generate fixed sub--varieties.

At any rate, one should eventually exclude non--perturbative instabilities of the type discussed in~\cite{hop}, and we leave to future work the analysis of all these issues. String corrections, which could drive the potential to a nearby stability domain, provide in our opinion an interesting alternative for these $AdS_3 \times S^7$ orientifold vacua.

\section{The (Generalized) $AdS_7 \times S^3$ Heterotic Vacua}\label{sec:heterotic_vacuum}

In these vacua there are a few novelties with respect to the orientifold case, since, which will have nonetheless some relevant effects. In the notation of Section~\ref{sec:model}, now $\beta=-1$, while
\beq
V \ = \ T\, e^{\,\frac{5}{2} \, \phi} \ .
\eeq
With this proviso, the equations of motion are again \eqref{eqsbeta} and \eqref{non_lag_Einstein}, but in this case the flux is in the internal sphere. Therefore eq.~\eqref{eqsbeta} becomes
\beq
\partial_i \left[ e^{\,\beta\,\phi} \, \sqrt{-g}\ H^{i j k}\right] \ = \ 0 \ ,
\eeq
which is solved by
\beq
H^{i j k} \ = \ \frac{\epsilon^{i j k}}{\sqrt{\gamma}}\ {\widetilde h} \  .
\eeq
Consequently, the scalar equation gives again
\beq
V_0^\prime \ = \ - \ \frac{\beta}{2} \ e^{\,\beta\,\phi_0} \left({\widetilde h} \right)^2 \ ,
\eeq
while the Einstein equations translate into
\bea
-\ \frac{1}{\beta}\ V_0^\prime &=& \frac{1}{2} \ e^{\,\beta\,\phi_0} \left({\widetilde h} \right)^2 \ = \ 2 \left( \frac{3}{R_{AdS}^2} \ + \ \frac{1}{R^2} \right) \ , \\
V_0 &=&  - \ \frac{36}{R_{AdS}^2} \ + \ \frac{4}{R^2} \ .
\eea
Notice that this solution demands that $\beta\, V_0' < 0$, and for later convenience we also define
\bea
\sigma_7 &=& 3 \ +\ \frac{R_{AdS}^2}{R^2} \ , \nonumber \\
\tau_7 &=& R_{AdS}\, V_0^{''} \ .
\eea
For these vacua $\sigma_7 \geq 3$, while the sign of $\tau_7$ is a priori arbitrary. Moreover,
\beq
R_{AdS}^2\,V_0 \ = \ 4 \left( \sigma_7 \ - \ 12 \right) \ ,
\eeq
which changes sign at $\sigma_7=12$, and finally for the \emph{torus--level} heterotic potential
\beq
\sigma_7 \ = \ 15 \ , \qquad \tau_7 \ = \ 75 \ .
\eeq

\subsection{Perturbations of the Generalized $AdS_7 \times S^3$ Heterotic Flux Vacua} \label{sec:perturbations73_gen}

In this case the {perturbed tensor equations} take the form
\bea
&\Box_{10}& \!\!\! b_{ij} \ - \ \nabla_i \Big( \nabla^M \, b_{M j} \Big) \ - \ \nabla_j \Big( \nabla^M \, b_{i M} \Big) \ - \ \frac{2}{R^2} \ b_{ij} \nonumber \\
&+& 4 \left( \frac{1}{R^2} \ + \ \frac{3}{R_{AdS}^2} \right)\ \sqrt{\gamma}\ \epsilon_{ijk}\, \Big( \beta\,\nabla^k\,\varphi \ - \ \nabla^\alpha \, {h_\alpha}^k \ - \ \frac{1}{2} \ \nabla^k\, \gamma \cdot h \ + \ \frac{1}{2} \ \nabla^k\, \lambda \cdot h \Big) \ = \ 0 \ , \nonumber \\
&\Box_{10}& \!\!\!  b_{i \mu} \ - \ \nabla_i \Big( \nabla^M \, b_{M \mu} \Big) \ - \ \nabla_\mu \Big( \nabla^M \, b_{i M} \Big) \ - \ \Big( \frac{2}{R^2} \ - \ \frac{6}{R_{AdS}^2} \Big)\,  b_{i\mu}  \label{beqs_73} \\
&+& 4 \left( \frac{1}{R^2} \ + \ \frac{3}{R_{AdS}^2} \right) \ \sqrt{\gamma}\ {\epsilon_{k l i}}\, \nabla^k\, {h^{l}}_\mu \ = \ 0 \ , \nonumber \\
&\Box_{10}& \!\!\!  b_{\mu \nu} \ - \ \nabla_\mu \Big( \nabla^M \, b_{M \nu} \Big) \ - \ \nabla_\nu \Big( \nabla^M \, b_{\mu M} \Big) \ + \ \frac{10}{R_{AdS}^2} \ b_{\mu\nu} \ = \ 0 \ , \nonumber
\eea
while the {perturbed dilaton equation} is now
\beq
\Box_{10}\, \varphi \ - \ V_0^{\prime\prime}\, \varphi \ - \ 2 \left( \frac{1}{R^2} \ + \ \frac{3}{R_{AdS}^2} \right) \left(\beta^2\, \varphi   \ - \ \beta\,\gamma \cdot h \right)
\ - \ \frac{\beta}{2}\ \frac{\epsilon^{\,i j k}}{\sqrt{\gamma}} \, \nabla_{i}\,b_{j k}\ = \ 0 \ .
\eeq
Finally, the {perturbed metric equations} that follow from
eq.~\eqref{non_lag_Einstein} rest on eq.~\eqref{pert_RMN} and read
\bea
\Box_{10} \, h_{i j} &-& \frac{2}{R^2} \, h_{i j} \ -\ \nabla_i \left(\nabla \cdot h\right)_{j}\ - \ \nabla_j \left(\nabla \cdot h\right)_{i}\ +\ \nabla_i\,\nabla_j \, (\lambda \cdot h + \gamma \cdot h) \nonumber \\ &+& \gamma_{i j}\Big[ \, \frac{5\,\beta}{2} \left(\frac{1}{R^2} + \frac{3}{R_{AdS}^2} \right) \varphi  + 3\,\gamma \cdot h \Big(\,\frac{1}{R^2} - \frac{3}{R_{AdS}^2} \Big) \, - \,  \frac{\epsilon^{\,k l m}}{4\,\sqrt{\gamma}} \, \nabla_{k}\,b_{l m}\Big] \nonumber \\
&+& \frac{1}{2\,\sqrt{\gamma}}\,\Big[ {\epsilon}^{\,k l m} \, \gamma_{i m}( \nabla_j\,b_{k l}+ \nabla_k \,b_{l j}+ \nabla_l\,b_{j k})\ + \ \left( i \leftrightarrow j \right)  \Big] \ = \ 0 \ ,\nonumber \\
\Box_{10} \, h_{i\mu} &-&  \Big( \frac{2}{R^2} + \frac{6}{R_{AdS}^2} \Big)h_{i\mu}  \ -\ \nabla_i \left(\nabla \cdot h\right)_{\mu}\ - \ \nabla_\mu \left(\nabla \cdot h\right)_{i}\ +\ \nabla_i\,\nabla_\mu \, (\lambda \cdot h + \gamma \cdot h)\nonumber \\ &+&  \,\frac{1}{2\,\sqrt{\gamma}} \ {\epsilon}^{\,k l m} \, \gamma_{i m} \, (\nabla_\mu\,b_{k l}+\nabla_k \,b_{l \mu}+\nabla_l \,b_{\mu k} )  \ = \ 0 \ , \label{tensor_het}\\
\Box_{10} \, h_{\mu \nu} &+& \frac{2}{R_{AdS}^2}\ h_{\mu \nu} \ -\ \nabla_\mu \left(\nabla \cdot h\right)_{\nu}\ - \ \nabla_\nu \left(\nabla \cdot h\right)_{\mu}\ +\ \nabla_\mu\,\nabla_\nu \, (\lambda \cdot h + \gamma \cdot h) \nonumber \\ &+&  \lambda_{\mu \nu}\left[  -\, \frac{2}{R_{AdS}^2} \ \lambda \cdot h \ - \
\left( \frac{1}{R^2} \,+\, \frac{3}{R_{AdS}^2}\right) \left(\frac{3}{2}\,\beta\,\varphi \,- \,\gamma \cdot h\right) \ - \ \frac{\epsilon^{\, i j k}\,\nabla_{i}\,b_{j k}}{4\,\sqrt{\gamma}}  \right] \ = \ 0 \ . \nonumber
\eea
In all cases, perturbations depend on the $AdS$ coordinates $x^\mu$ and on the sphere coordinates $y^i$, and will be expanded in the corresponding spherical harmonics, whose structure is briefly reviewed in Appendix~\ref{sec:app2}, proceeding as in Section~\ref{sec:perturbations37_gen}.

\subsection{Tensor and Vector Perturbations} \label{tens_vec_73}

To begin with, let us notice that tensor perturbations, which result from transverse traceless $h_{\mu\nu}$ when all other perturbations vanish, satisfy in these $AdS_7$ vacua the equation
\beq
\Box \, h_{\mu\nu} \ - \ \frac{\ell(\ell+2)}{R^2}\ h_{\mu\nu} \ +\  \frac{2}{R_{AdS}^2} \, h_{\mu\nu} \ = \ 0 \ \qquad ( \ell \geq 0) \ .
\eeq
For $\ell=0$, this describes a massless graviton field, which is accompanied by a tower of Kaluza--Klein states for higher $\ell$. Hence, there are \emph{no instabilities in this sector}.

There are massive scalar excitations resulting from the traceless part of $h_{ij}$ that is also divergence free, which is a tensor with respect to the internal rotation group. They satisfy
\beq
\Box \, h_{i j} \ - \ \frac{\ell(\ell+2)(\sigma_7-3)}{R_{AdS}^2}\ h_{ij} \ = \ 0 \qquad ( \ell \geq 2)\ ,
\eeq
so that the results in Appendix~\ref{sec:app2} imply that \emph{no instabilities are present also in this sector}.
There are also \emph{no unstable modes} from transverse $b_{\mu\nu}$ excitations, which satisfy,
\beq
\Box\,b_{\mu \nu} \ + \ \frac{[10 \,-\, \ell(\ell+2)(\sigma_7-3)]}{R_{AdS}^2} \ b_{\mu\nu} \ = \ 0 \qquad ( \ell \geq 0)\ ,
\eeq
so that the lowest ones, corresponding to $\ell=0$, are massless.

Vector perturbations are more involved, since there are mixings between $h_{i\mu}$ and $b_{i \mu}$. In order to study them, one can proceed in analogy with Section \ref{sec:tens_vec_37}, letting~\footnote{{ The function $F_\mu^q$ is not uniquely defined, and the resulting freedom makes it possible to eliminate source terms that could in principle affect some of the following equations.}}
\beq
b_{i\mu} \ = \ \sqrt{\gamma}\, \epsilon_{\,i p q}\, \nabla^p\, {F_\mu}^q \ ,
\eeq
which is transverse in internal space. The resulting system reads
\bea
R_{AdS}^2\,\Box\, F_{\mu i} &+& [ 6 - (\ell+1)^2(\sigma_7-3)]F_{\mu i} \ + \ 4\,\sigma_7 \, h_{\mu i} \ = \ 0 \ ,\nonumber \\
R_{AdS}^2\,\Box\, h_{\mu i} &+& (\ell+1)^2(\sigma_7-3) F_{\mu i} \ - \ [ 6 +(\ell+1)^2(\sigma_7-3)]h_{\mu i} \ =  0 \ ,
\eea
and the eigenvalues are
\beq
(\ell\,+\,1)^2(\sigma_7\,-\,3) \ \pm \ 2\, \sqrt{\sigma_7(\sigma_7\,-\,3)(\ell\,+\,1)^2\ +\ 9} \ .
\eeq

In order to refer to the BF bound in Appendix~\ref{sec:app3} one should add 6 to these expressions and compare the result with $-4$. All in all, \emph{there are no modes below the BF bound in this sector.} The vector modes are massive for $\ell>1$ in the region $\sigma_7>3$, while they become massless for $\ell=1$ and all allowed values of $\sigma_7 > 3$, and for all values of $\ell$ in the singular limit $\sigma_7=3$, which would correspond to a three--sphere of infinite radius. All in all, for $\ell=1$ there are 6 massless vectors arising from one of the two eigenvalues above in the heterotic vacuum. According to Appendix \ref{sec:app2}, they build up an antisymmetric tensor in internal vector indices, and therefore an adjoint multiplet of $SO(4)$ vectors. The counting is consistent with Kaluza--Klein theory and with the internal symmetry of $S^3$, although the massless vectors originate here from mixed contributions of the metric and the two--form field.

\subsection{Scalar Perturbations} \label{sec:scalar_pert_73}

Proceeding as in the preceding sections, let us now concentrate on scalar perturbations starting from
\beq
g_{MN} = g^{(0)}_{MN} + h_{MN} \ , \quad \phi = \phi_0 + \varphi \ , \quad B_{MN}=B_{0\,MN}\ +\ e^{\,-\,\beta\,\phi_0} \, \frac{b_{MN}}{{\widetilde h}} \ .
\eeq
One can describe scalar perturbations letting
\beq
b_{\mu\nu} \ = \ 0 \ , \qquad b_{\mu i} \ = \ 0 \ ,
\qquad
b_{i j} \ = \ \sqrt{\gamma} \ \epsilon_{i j k} \, \nabla^k \, B \ ,
\eeq
a choice that also satisfies
\beq
\nabla^M\,b_{MN} \ = \ 0 \ .
\eeq
In addition, let us parametrize scalar metric perturbations as
\beq
h_{\mu\nu} \ = \ \gamma_{\mu\nu}\, A \ , \qquad h_{\mu i} \ = \ R_{AdS}^2\, \nabla_\mu\, \nabla_i\,D \ , \qquad h_{ij} \ = \ \gamma_{ij}\,C \ ,
\eeq
along the lines of what we did in the preceding sections.

In this fashion, the first of eqs.~\eqref{beqs_73} turns into
\beq
\left(\Box \  - \ \frac{\ell(\ell+2)}{R^2}\right)\, B \ - \ \frac{4\,\sigma_7}{R_{AdS}^2}\, \left( -\,\beta\,\varphi \,+\, R_{AdS}^2\,\Box\,D  \,+\, \frac{1}{2}\left( 3\,C\,-\, 7\,A\right)\right) \ = \ 0 \ ,
\eeq
which is now present only for $l \neq 0$, while the complete scalar equation reads
\beq
\Box \, \varphi \ + \ \frac{\ell(\ell+2)}{R^2}\ (\beta\,B \,-\, \varphi) \ - \ \frac{2\,\sigma_7}{R_{AdS}^2} \left(\beta^2\,\varphi \,+\,3\,\beta\,C \right) \ - \ V_0^{\prime\prime}\, \varphi  \ = \ 0 \ .
\eeq
\begin{figure}[ht]
\begin{center}
\includegraphics[width=90mm]{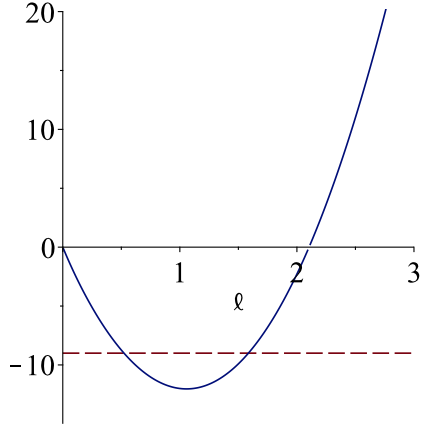}
\vspace*{-5.5truecm}
\end{center}
\caption{\small Violation of the BF bound in heterotic flux vacua. The dangerous eigenvalue is displayed in units of $\frac{1}{R_{AdS}^2}$ and the BF bound is -9 in this case.}
\label{fig:bad_eigenvalue_het}
\end{figure}

For scalar perturbations one can thus obtain
\bea
\Box\, A &-& \left(\frac{12}{R_{AdS}^2} \ + \ \frac{\ell(\ell+2)}{R^2}\right)A \ + \ \frac{3\,\sigma_7}{R_{AdS}^2} \left( -\, \frac{\beta}{2} \ \varphi \ + \ C  \right) \ + \ \frac{\ell(\ell+2)}{2\, R^2}\ B \ = \ 0 \ , \nonumber \\
5\, A &+& 3\, C \ + \ 2\,{\ell(\ell+2)}\,(\sigma_7-3)\,D \ = \ 0 \
,\nonumber \\
6\,A &+& 2\, C \ + \ B \ - \ {12}\, D \ = \ 0 \ ,  \\
7\,A &+& C \ - \ 2\, R_{AdS}^2\,\Box\, D \ = \ 0 \ , \nonumber \\
\Box\,C &-& \frac{\ell(\ell+2)}{R^2}\ C \ - \ \frac{5\,\sigma_7+12}{R_{AdS}^2}\,C \ + \ \frac{5\,\beta\,\sigma_7}{2\,R_{AdS}^2} \ \varphi  \ - \ \frac{3 \ell(\ell+2)}{2\, R^2} \ B \ = \ 0 \ , \nonumber\\
\Box \, \varphi &+& \frac{\ell(\ell+2)}{R^2}\ (\beta\,B \,-\, \varphi) \ - \ \frac{2\,\sigma_7}{R_{AdS}^2} \left(\beta^2\,\varphi \,-\,3\,\beta\,C \right) \ - \ V_0^{\prime\prime}\, \varphi  \ = \ 0 \ , \nonumber \\
\Box\,B &-& \frac{\ell(\ell+2)}{R^2}\, B \ - \ \frac{4\,\sigma_7}{R_{AdS}^2}\, \left(-\,\beta\, \varphi \,+\, 2\,C\right) \ = \ 0 \ , \nonumber
\eea
where we have eliminated $\Box\,D$ from the tensor equation. Although there are seven equations for five unknowns, one can verify that the system is consistent. All in all, one can thus work with $(C,\varphi,B)$, restricting the attention to
\bea
R_{AdS}^2\,\Box\,C &-& \Big[{L_7}(\sigma_7-3)\ +\ 5\,\sigma_7 \ + \ 12\Big]\,C \ + \ \frac{5\,\beta\,\sigma_7}{2} \ \varphi \ - \ \frac{3\,L_7}{2}\ (\sigma_7 -3) \ B \ = \ 0 \ , \nonumber\\
R_{AdS}^2\,\Box\,\varphi &+& 6\,\beta\,\sigma_7\,C \ - \ \Big[L_7(\sigma_7-3) + 2\,\beta^2\,\sigma_7\ + \ \tau_7 \Big]\,\varphi \ + \ {L_7}\, \beta\,(\sigma_7 -3) \ B \ = \ 0 \ , \\
R_{AdS}^2\,\Box\,B &-& 8\,\sigma_7\,C \ + \  4\,\beta\,\sigma_7\,\varphi \ - \ {L_7}\, (\sigma_7 -3) \ B \ = \ 0 \ , \nonumber
\eea
which are here expressed in terms of the two variables $\sigma_7$ and $\tau_7$ of Section \ref{sec:heterotic_vacuum}, to then determine $A$ and $D$ algebraically, and where for brevity we have let
\beq
L_7 \ = \ \ell(\ell+2) \ .
\eeq

To begin with, let us notice that for $\ell=0$ $B$ decouples, so that the system reduces to
\bea
R_{AdS}^2\,\Box\,C &-& (5\,\sigma_7 \ + \ 12\Big)\,C \ + \ \frac{5\,\beta\,\sigma_7}{2} \ \varphi  \ = \ 0 \ , \nonumber\\
R_{AdS}^2\,\Box\,\varphi &+& 6\,\beta\,\sigma_7\,C \ - \ (2\,\beta^2\,\sigma_7\ + \ \tau_7)\,\varphi  \ = \ 0 \ .
\eea
\begin{figure}[ht]
\begin{center}
\includegraphics[width=110mm]{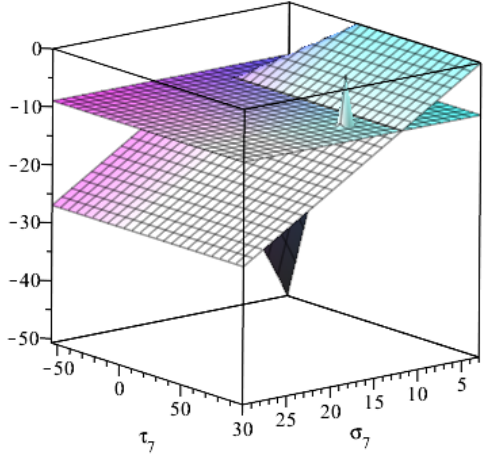}
\vspace*{-6.5truecm}
\end{center}
\caption{\small Comparison between the lowest eigenvalue of $R_{AdS}^2\,{\cal M}^2$ and the BF bound, which is -9 in this case. There are regions of stability for values of $\sigma_7$ close to 3, which correspond to $\frac{R^2}{R_{AdS}^2}>9$, and to negative values of $V_0$. The example displayed here refers to $\beta=-1$ and $\ell=1$, which corresponds to the minimum in the heterotic case of fig.~\ref{fig:bad_eigenvalue_het}, and the peak identifies the point corresponding to the tree--level values $\sigma_7=15$, $\tau_7=75$.}
\label{fig:bad_eigenvalue_gen_het}
\end{figure}
In terms of the $\sigma_7$ and $\tau_7$ variables of Section~\ref{sec:heterotic_vacuum}, the mass matrix thus reads
\beq
R_{AdS}^2\, {\cal M}^2 \ = \ \left(
\begin {array}{cc}
5\,\sigma_7+12 & -\ \frac{5}{2}\ \beta\,\sigma_7 \\  -\  6\,\beta\,\sigma_7 & \ \ 2\,\beta^2\,\sigma_7 \ +  \ \tau_7
\end {array}
\right)     \ ,
\eeq
and its eigenvalues are
\beq
{\beta}^{2}\sigma_7+\frac{5\,\sigma_7}{2}+\frac{\tau_7}{2}+6\, \pm\, \frac{1}{2}\, \sqrt{\left( 4\,{\beta}^{4}+40\,{\beta}^{2
}+25 \right) {\sigma_7}^{2}+4 \left( \tau_7-12 \right)  \left( {\beta}^{2}-5/2
 \right) \sigma_7+ \left( \tau_7-12 \right) ^{2}
}
 \ .
\eeq
They are both positive for the tree--level effective Lagrangian,
and for the heterotic potential, for which
\beq
\sigma_7 \ = \ 15 \ , \qquad \tau_7 \ = \ 75 \ , \qquad \beta \ = \ - \ 1 \ ,
\eeq
they read
\beq
\lambda_{1,2} \ = \frac{24}{R_{AdS}^2} \left(4 \ \pm \ \sqrt{6} \right) \ .
\eeq

We can now move on to the $\ell \neq 0$ case, where the three functions $(C,\phi,B)$ all contribute, so that one is led to the mass matrix
\beq
R_{AdS}^2\,{\cal M}^2 \ = \ L_7 (\sigma_7 -3) {\mathbf 1}_3 \ + \  \left( \begin{array}{ccc} 5\,\sigma_7+12 & -\ \frac{5\,\beta}{2}\ \sigma_7 & \frac{3}{2} \, L_7 (\sigma_7 -3) \\- \ 6\,\beta\,\sigma_7 & \ \ 2\,\beta^2\,\sigma_7+ \tau_7 &\  - \ \beta\, L_7(\sigma_7 -3) \\ 8\,\sigma_7 & -\ 4\,\beta\,\sigma_7 & 0  \end{array}\right)\ .
\eeq
\begin{figure}[ht]
\begin{center}
\includegraphics[width=110mm]{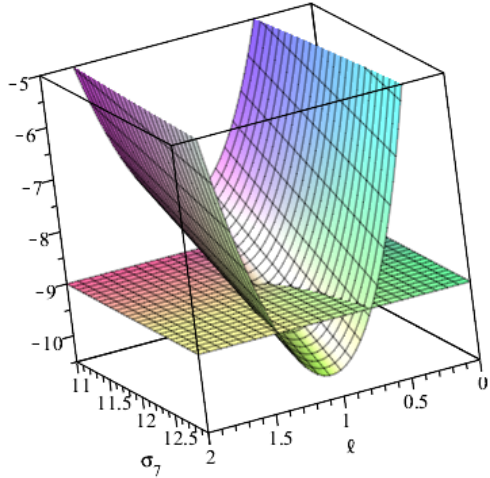}
\vspace*{-6truecm}
\end{center}
\caption{\small A different view. Comparison between the lowest eigenvalue of $R_{AdS}^2\,{\cal M}^2$ and the BF bound, which is -9 in this case, as a function of $\ell$ and $\sigma_7$, for $\tau_7=75$. There are regions of stability for values of $\sigma_3$ below 12, which correspond to relatively large values for the ratio $\frac{R^2}{R_{AdS}^2}$ and to negative values of $V_0$.}
\label{fig:bad_eigenvalue_het_different}
\end{figure}
In most of the parameter space, two eigenvalues are not problematic, but there is one bad eigenvalue in the tree--level heterotic potential, which corresponds to $\sigma_7=15$ and $\tau_7=75$. It obtains for $\ell=1$ and $k=0$ from
\beq
4\left[ 16 \,+\, 3\, L_7 \,-\, 4\, \sqrt{34 \,+\,15\,L_7}\, \cos\left( \frac{\delta - 2\pi\,k}{3}\right)\right]  \qquad (k=0,1,2) \ ,
\eeq
where
\beq
\delta \ = \ {\mathrm{Arg}}\left( 152\,-\,45 L_7\,+\, 3\,i \,\sqrt{3(5 L_7\,+\,3)[(5 L_7\,+\,14)^2\,+\,4]}\right) \ .
\eeq

Still, there is again a stability region \emph{for values of $\sigma_7$ that are close to $12$, for negative $V_0$, and typically for positive $\tau_7$, {\it i.e.} for potentials that are convex close to the vacuum configuration.}. These results are displayed in figs.~\ref{fig:bad_eigenvalue_gen_het} and \ref{fig:bad_eigenvalue_het_different}.

In this case one could eliminate the bad eigenvalue by a $\mathbb{Z}_2$ antipodal projection in the internal sphere $S^3$, which can be identified with the $SU(2)$ group manifold. This operation has no fixed points,
and reduces the internal space to the $SO(3)$ group manifold, without affecting the massless vectors with $\ell=1$ that we have identified. Alternatively, one could resort to the symmetry group of the sphere related to the action on unit quaternions that we have described for the orientifold vacua. The resulting settings, however, resonate again with the discussion in~\cite{hop}, and non--perturbative instabilities are in principle relevant to the story, so that we leave an analysis of this problem to future work. String corrections would also deserve a closer look, since they could drive the potential to a nearby stability domain, providing an interesting alternative for these $AdS_7 \times S^3$ heterotic vacua.

\section{The 9D Dudas--Mourad Non--Supersymmetric Vacua}\label{sec:dm_vacua}

Our starting point is now provided by the classical solutions of~\cite{dm_9Dsolution} for the ten--dimensional Sugimoto orientifold~\cite{sugimoto}, which in the Einstein frame read
\begin{eqnarray}
\phi &=& \frac{3}{4}\ \alpha_O y^2\ + \ \frac{1}{3}\ \ln \left|{\alpha_O} y^2\right| \ + \ \Phi_0 \ , \nonumber \\
ds_O^2 &=& \left|{\alpha_O} \,y^2\right|^\frac{1}{18} e^{- \alpha_O \frac{y^2}{8}}\  \eta_{\mu\nu}dx^\mu dx^\nu
\ + \ e^{-\frac{3}{2}\, \Phi_0}\,\left|{\alpha_O} y^2\right|^{-\frac{1}{2}} e^{- \frac{9}{8}\, \alpha_O y^2} dy^2 \  \label{s1o}\ , \label{9D_orientifold_solution}
\end{eqnarray}
where the absolute values would not be needed here, since $y \in (0,\infty)$, but this notation will be convenient in the following section. The corresponding Einstein--frame expressions for the $SO(16) \times SO(16)$ {heterotic model} of~\cite{so16xso16} read
\begin{eqnarray}
\phi &=& \frac{1}{2}\ \ln \sinh\left|\sqrt{\alpha_H} y \right| \ + \ 2\, \ln \cosh\left|\sqrt{\alpha_H} y\right| \ + \ \Phi_0 \ , \nonumber \\
ds_H^2 &=& \left|\sinh\left(\sqrt{\alpha_H} y \right)\right|^\frac{1}{12}\,\left|\cosh\left(\sqrt{\alpha_H} y \right)\right|^{-\frac{1}{3}} \  \eta_{\mu\nu}dx^\mu dx^\nu
\nonumber \\ &+& e^{-\frac{5}{2}\, \Phi_0}\,\left|\sinh\left(\sqrt{\alpha_H} y \right)\right|^{-\frac{5}{4}}\,\left|\cosh\left(\sqrt{\alpha_H} y \right)\right|^{-5} \,
 dy^2 \  \label{s1h}\ . \label{9D_heterotic_solution}
\end{eqnarray}

These two vacua have nine--dimensional Poincar\'e symmetry, while their internal spaces are actually intervals of finite length in the metrics of eqs.~\eqref{9D_orientifold_solution} and \eqref{9D_heterotic_solution}. It is convenient to recast the two solutions in terms of conformally flat metrics, so that one is led to consider expressions of the type
\begin{eqnarray}
&& ds^2 \ = \ e^{2\,\Omega(z)} \left( \eta_{\mu\nu}\, dx^\mu\,dx^\nu \ + \ dz^2 \right) \ , \nonumber \\
&& \phi \ = \ \phi(z)
\ . \label{9D_conformal}
\end{eqnarray}
In detail, in the orientifold case the coordinate $z$ is obtained integrating the relation
\beq
dz \ = \ \left| \alpha_O\, y^2 \right|^{-\,\frac{5}{18}}\, e^{\,-\,\frac{3}{4}\, \Phi_0} \, e^{-\frac{\alpha_O}{2}\, y^2}\ d y \ ,
\eeq
while
\beq
e^{2\,\Omega(z)} \ = \ |\sqrt{\alpha_O} \,y|^\frac{1}{9} e^{- \frac{\alpha_O}{8} y^2} \ . \label{omega_orientifold}
\eeq
On the other hand, in the heterotic case
\beq
dz \ = \ e^{-\frac{5}{4}\, \Phi_0} \left|\sinh\left(\sqrt{\alpha_H} y \right)\right|^{-\frac{2}{3}}\,\left|\cosh\left(\sqrt{\alpha_H} y \right)\right|^{-\frac{7}{3}} \ d y \ ,
\eeq
and
\beq
e^{2\,\Omega(z)} \ = \ \left|\sinh\left(\sqrt{\alpha_H} y \right)\right|^\frac{1}{12}\,\left|\cosh\left(\sqrt{\alpha_H} y \right)\right|^{-\frac{1}{3}} \ . \label{omega_heterotic}
\eeq

Notice that one is confronted with an interval whose finite length is proportional to $\frac{1}{\sqrt{\alpha_{O,H}}}$ in the two cases, but which hosts a pair of metric singularities at its two ends $y=0$ and $y=+ \infty$, with a string coupling that is weak at the former and strong at the latter. Moreover, the parameters $\alpha_O$ and $\alpha_H$ are proportional to the residual tension, and therefore the supersymmetric case corresponds to the singular limit as they tend to zero, where the internal length diverges.

\subsection{Perturbations of the 9D Vacua}\label{sec:pert_9D_vacuum}

The equations of interest are now
\bea
\Box\phi  &-&  \frac{\partial V}{\partial \phi} \ = \ 0 \ , \nonumber \\
R_{MN} &+& \frac{1}{2} \, \pr_M\phi\, \pr_N\phi\ + \ \frac{1}{8}\ g_{MN}\, V(\phi) \ = \ 0 \label{96_field_eqs} \ ,
\eea
and the corresponding perturbed fields take the form
\begin{eqnarray}
&& ds^2 \ = \ e^{2\,\Omega(z)} \Big(\eta_{MN}\,+\,h_{MN}(x,z) \Big) dx^M\,dx^N  \ , \nonumber \\
&& \phi \ = \ \phi(z) \ + \ \varphi(x,z)
\ . \label{9D_solution_conformal}
\end{eqnarray}
As a result, the perturbed Ricci curvature can be extracted from
\bea
R_{MN} &=& 8\, \nabla_M\, \nabla_N\, \Omega \ + \ (\eta_{MN}+h_{MN})\, \nabla^A\,\nabla_A\,\Omega \nonumber \\ &-& 8\ \Big( \nabla_M\, \Omega\,\nabla_N\, \Omega \ - \ (\eta_{MN}+h_{MN}) \,\nabla^A\,\Omega \,\nabla_A\,\Omega \Big) \nonumber \\
&+& \frac{1}{2} \, \Big[(\Box\,+\, \partial_z^2)\, h_{MN} \ -\ \nabla_M \left(\nabla \cdot h\right)_{N}\ - \ \nabla_N \left(\nabla \cdot h\right)_{M}\ +\ \nabla_M\,\nabla_N \, h_L{}^L \Big] \ ,
\eea
an expression where covariant derivatives do not involve $\Omega$, and thus refer to $\eta_{MN}\,+\,h_{MN}$, which is also used to raise and lower indices. Up to first order the metric equations of motion thus read
\bea
R_{MN} &+& \frac{1}{2}  \pr_M\phi \,\pr_N\phi\ +\ \frac{1}{2}\,  \pr_M\varphi \,\pr_N\phi\
+\frac{1}{2} \pr_M\phi \,\pr_N\varphi\nonumber \\
&+& \frac{1}{8}\ e^{2\Omega}\Big[(\eta_{MN}+h_{MN}) V+\eta_{MN}V_\phi\,\varphi\Big] \ = \ 0\ , \label{gravity_firstorder}
\eea
and combining this result and the scalar equation in~\eqref{96_field_eqs} yields the unperturbed equations of motion
\bea
\Omega''\ +\ 8\,(\Omega')^2\ +\ \frac{e^{2\Omega}}{8}\,V(\phi)&=&0\ ,\\
9\,\Omega''\ +\ \frac{e^{2\Omega}}{8}\,V(\phi) \ +\ \frac{1}{2}\,(\phi')^2&=&0\ ,\\
\phi''\ +\ 8\,\Omega'\phi'\ -\ e^{2\Omega}\,V_\phi(\phi) &=&0\ , \label{eqfi}
\eea
where $V$ and $V_\phi$ denote the potential and its derivative computed on the classical vacuum.
Notice that the first two equations can be equivalently recast in the form
\bea
72(\Omega')^2\ -\ \frac{1}{2}\,(\phi')^2\ +\ e^{2\Omega}
\,V(\phi)&=&0\ ,\\
8\big[\Omega''\ -\ (\Omega')^2\big]\ +\ \frac{1}{2}\, (\phi')^2&=&0\ ,
\eea
and that the equation for $\phi$ is a consequence of these.

All in all, \eqref{gravity_firstorder} finally leads to
\bea
-\frac{1}{8}\ e^{2\Omega}\,\eta_{\mu\nu}V_\phi\,\varphi &=& -4(\partial_\mu h_{\nu 9}+\partial_\nu h_{\mu 9}- h_{\mu\nu}')\Omega' \nonumber\\&-&\eta_{\mu\nu} \Big[h_{99}
[\Omega''+8(\Omega')^2]+(\partial_\alpha h^{\alpha9}-\frac{1}{2} [{{h'}_\alpha{}^{\alpha}}-h_{99}'])\Omega'\Big] \nonumber\\
&+& \frac{1}{2} \Big[\Box\,h_{\mu\nu}+ h_{\mu\nu}'' \ - \partial_\mu \left(
\partial_\alpha  h^{\alpha}{}_\nu+ h_{9\nu}'\right)\nonumber \\ &-&  \partial_{\nu} \left(
\partial_\alpha
h^{\alpha}{}_\mu+{h'}_{9\mu}\right)+ \partial_\mu \partial_\nu ( h_\alpha{}^\alpha+h_{99}) \Big]\ , \\
-\frac{1}{2}  \pr_\mu\varphi \,\phi' &=& -4\, \partial_\mu h_{99}\, \Omega'  \   \nonumber \\
&+& \frac{1}{2}  \Big[\Box\,h_{\mu 9}  - \partial_\mu
\partial_\alpha h^\alpha{}_9 \ - \
\partial_\alpha {h'}^\alpha{}_\mu \ +\partial_\mu {h'}_\alpha{}^\alpha \Big]\ ,  \\
 - \varphi'\, \phi'\
- \frac{1}{8}\ e^{2\Omega}\Big[h_{99} V+V_\phi\,\varphi\Big] &=& -4{h'}_{99}\Omega' \ - \Big[\partial_\alpha h^{\alpha9}-\frac{1}{2}[{h'}_\alpha{}^\alpha-{h'}_{99}]\Big]\Omega'
 \nonumber \\
&+& \frac{1}{2} \, \Big[\Box \, h_{99} \ -2\,
\partial_\alpha {h'}^\alpha{}_9 + {h''}_\alpha{}^\alpha  \Big] \ .
\eea

\subsection{Scalar Perturbations} \label{sec:scalar9d}
For scalar perturbations
\bea
h_{\mu\nu} \ = \ \eta_{\mu\nu}\, e^{ip.x}A(z) \ , \quad h_{\mu9} \ = \ i\,p_\mu\,D(z)e^{ip.x} \ , \quad
h_{99}=e^{ip.x}C(z)\ ,
\eea
with $p^\mu p^\nu\eta_{\mu\nu}=-m^2$ and $p.x=p^\mu x^\nu\eta_{\mu\nu}$, so that the Einstein equations become
\bea
-\frac{1}{8}\ e^{2\Omega}\Big[A V+V_{\phi}\varphi\Big] \eta_{\mu\nu}&=& 4(2\,p_\mu\,p_\nu D + A'\eta_{\mu\nu})\Omega'+
 \eta_{\mu\nu}A[\Omega''+8(\Omega')^2]\nonumber\\&-&\eta_{\mu\nu} \Big[ C
[\Omega''+8(\Omega')^2]+(m^2 D-\frac{1}{2}[9A'-C'])\Omega'\Big] \nonumber\\
&+& \frac{1}{2} \Big[\eta_{\mu\nu}(A'' + m^2 A) \ - \ p_\mu \, p_\nu ( 7A+C-2 D') \Big],\ \ \\
-\frac{1}{2}  \varphi \,\phi'\
- \frac{1}{8}\ e^{2\Omega} D V &=& -4\, C\, \Omega' \ + \ D\, [\Omega''+8(\Omega')^2] \  +4 A'   \\
- \varphi' \,\phi'\
- \frac{1}{8}\ e^{2\Omega}\Big[C V+V_{\phi}\varphi\Big] &=& -4 C' \Omega' \ - \Big[m^2 D-\frac{1}{2}(9A'-C')\Big]\Omega'
 \nonumber \\
&+& \frac{1}{2} \Big[\, m^2( \, C \ -2\
D' )+ 9 A''  \Big] \ ,
\eea
while the perturbed scalar equation reads
\beq
\Box\,\varphi + \varphi'' + 8 \Omega'\,\varphi'+ \phi'\left[ \frac{1}{2} (h_\mu^\mu)'- \frac{1}{2} (h_9^9)' -
\partial_\mu h^{\mu 9} - 8\Omega' h_{99}\right] - \phi'' h_{99} - e^{2\Omega} V_{\phi\phi}\, \varphi\ = \ 0 \ .
\eeq
Substituting the preceding expressions and using the background equations now gives
\beq
m^2 \varphi + \varphi'' + 8 \Omega'\,\varphi'+ \phi'\left[ \frac{9}{2} A'- \frac{1}{2} C' - m^2\,D \right] - C\,e^{2\Omega} V_\phi - e^{2\Omega} V_{\phi\phi} \, \varphi\ = \ 0 \ , \label{eqphi_9d}
\eeq
and altogether the four scalars $A$, $C$, $D$ and $\phi$ obey the linearized equations
\bea
-\,\frac{1}{8}\ e^{2\Omega}V_{\phi}\,\varphi &=&
  -\ \Big[m^2 D-\frac{1}{2}\left(17A'-C'\right)\Big]\,\Omega'
\nonumber \\&+& \frac{1}{2} (m^2\, A \ + \ A'') \ - \  C
\Big[\Omega''+8(\Omega')^2\Big] \nonumber \ , \\
  7\,A\ +\ C\ -\ 2\,D' \ -\ 16\, D\,\Omega' &=& 0 \ , \nonumber \\
 4\, C\, \Omega'  \ -   \ 4\, A' \ -\ \frac{1}{2}\,  \varphi \,\phi'\
 &=& 0  \label{eqsmetric_9d} \ , \\
- \ \varphi'\, \phi' \
- \ \frac{1}{8}\ e^{2\Omega}\Big[C\, V\ +\ V_{\phi}\,\varphi\Big] &=&  - \Big[m^2 D-\frac{9}{2} \,(A'\,-\,C')\Big]\Omega'
 \nonumber \\
&+& \frac{1}{2} \Big[\, m^2( \, C \ -2\,
D' )+ 9\, A''  \Big] \ . \nonumber
\eea
Notice that some of the metric equations, the second one and the third one above, are constraints, and that there is actually another constraint that obtains combining the first and the last so as to remove $A''$. Moreover, eq.~\eqref{eqphi_9d} is a consequence of these.

The system, however, has a residual local gauge symmetry, a diffeomorphism of the type
\beq
z' \ = \ z \ + \ \epsilon(x^\mu,z) \ ,
\eeq
which is available in the presence of a single internal dimension and implies
\beq
dz \ = \ dz'\left(1 \,-\, \frac{d\epsilon}{dz'} \right) \ - \ dx^\mu \partial_\mu \, \epsilon \ .
\eeq
Taking into account the original form of the metric in eq.~\eqref{9D_solution_conformal}, which in terms of our perturbations reads
\beq
ds^2 \ = \ e^{2\Omega(z)} \left[ (1+ A) \,dx^\mu \, dx_\mu  \ + \ 2\, dx^\mu \,dz \,\partial_\mu D \ + \ (1+C)\, dz^2 \right] \ ,
\eeq
one can thus identify the transformations
\bea
A &\rightarrow& A \ - \ 2\, \Omega'\, \epsilon \ , \nonumber \\
C &\rightarrow& C \ - \ 2\, \Omega'\, \epsilon \ - \ 2\, \epsilon' \ , \nonumber \\
D &\rightarrow& D \ - \ \epsilon \ , \nonumber \\
\varphi &\rightarrow& - \ \phi'\, \epsilon \ .
\eea
Notice that $D$ behaves as a St\"uckelberg field, and can be gauged away, leaving only one scalar degree of freedom after taking into account the constraints, as expected from Kaluza--Klein theory.

After gauging away $D$ the second of eqs.~\eqref{eqsmetric_9d} implies that
\beq
C \ = \ - \,7\,A \ ,
\eeq
while the third of eqs.~\eqref{eqsmetric_9d} implies that
\beq
\varphi \ = \ - \ \frac{8}{\phi'} \left(A'\ + \ 7\,A\,\Omega'\right) \ . \label{phi_A}
\eeq
Substituting these expressions in the first of eqs.~\eqref{eqsmetric_9d} finally leads to a second--order eigenvalue equation for $m^2$:
\beq
A'' + A'\left(24\, \Omega'\ - \ \frac{2}{\phi'} \ e^{2\Omega}\, V_{\phi} \right) \ + \
A\left(m^2 \ - \ \frac{7}{4} \ e^{2\Omega}\, V \ - \ 14 \, e^{2\Omega}\, \Omega'\, \frac{V_\phi}{\phi'} \right) \ = \ 0 \ . \label{second_order_9d}
\eeq
There is nothing else, since differentiating the third of eqs.~\eqref{eqsmetric_9d} and using the background equations gives
\beq
\varphi'\, \phi'\ = \ - \ 8\, A'' \ - \ 120\, A'\,\Omega'\ + \ 8\, e^{2\Omega}\, \frac{V_\phi}{\phi'}\ A'\ + \ 56  \,e^{2\Omega}\, \frac{V_\phi}{\phi'} \ \Omega'\, A \ + \ 7 \, e^{2\Omega}\, V\, A \ ,
\eeq
Taking this result into account, one can verify that the last of eqs.~\eqref{eqsmetric_9d} also leads to \eqref{second_order_9d}, whose properties we now turn to discuss.

The issue at stake is again the stability of the solution, which in this case reflects itself in the sign of $m^2$: a negative value would signal a tachyonic instability in the nine--dimensional Minkowski space. Changing parameters in the potential is more complicated in this case, since the background values depend on $z$, but nonetheless we can show that the solution corresponding the lowest--order level potentials is stable, in both the orientifold and heterotic models.
To this end, let us recall that a generic second--order equation of the type
\beq
f''(z) \ + \ a(z) \, f'(z) \ + \ \left[m^2 \ - \ b(z)\right]\, f(z) \ = \ 0 \ , \label{eq_schro}
\eeq
can be turned into a Schr\"odinger--like form via the transformation
\beq
f(z) \ = \ \Psi(z) \, e^{\,- \,\frac{1}{2} \int a\, dz} \ .
\eeq
One is thus led to
\beq
\frac{d^2 \Psi}{dz^2} \ + \ \left(m^2 \ - \ b \ - \ \frac{a'}{2} \ - \ \frac{a^2}{4} \right) \Psi \ = \ 0 \ , \label{schroed}
\eeq
and tracing the preceding steps one can see that $\Psi$ is in $L^2$. Eq.~\eqref{schroed} can be conveniently discussed connecting it to a more familiar problem of the type
\beq
m^2\, \Psi \ = \  {\cal H}\,\Psi  \ ,
\eeq
where
\beq
{\cal H} \ = \ b \ + \ {\cal A}^\dagger\, {\cal A} \ ,
\eeq
with
\beq
{\mathcal A}  \ = \  - \ \frac{d}{dz} \ + \ \frac{a}{2} \ , \qquad
 {\cal A}^\dagger \ = \  \frac{d}{dz} \ + \ \frac{a}{2} \ .
\eeq
Once these relations are supplemented with Neumann or Dirichlet conditions at each end in $z$, one can conclude that in all these cases
\beq
{\cal A}^\dagger\, {\cal A}  \ \geq \ 0 \ .
\eeq

All in all, positive values of $b$ then imply positive values of $m^2$, and this condition is indeed realized for the 9D orientifold vacuum, since
\beq
b \ = \ \frac{7}{4} \, e^{2\Omega}\left( V \ + \ 8 \, \Omega'\, \frac{V_\phi}{\phi'}\right)\ ,
\eeq
and the corresponding $V \sim e^{\frac{3}{2}\, \phi}$, so that
\beq
b \ = \ \frac{7}{4} \, e^{2\Omega}\, V \left( 1 \ + \ 12 \, \frac{\Omega'}{\phi'}\right) \ .
\eeq
The ratio of derivatives can be simply computed in terms of the $y$ coordinate using eqs.~\eqref{9D_orientifold_solution} and \eqref{omega_orientifold}, which yields the manifestly non--negative expression
\beq
b \ = \ {7} \, e^{2\Omega}\, V \, \frac{1}{1 \ + \ \frac{9}{4}\, \alpha_O\, y^2} \ .
\eeq
To reiterate, \emph{the 9d orientifold vacuum is a perturbatively stable solution of the Einstein--dilaton system} for all allowed choices of boundary conditions at the ends of the interval.

In the heterotic case $V \sim e^{\frac{5}{2}\, \phi}$, and
\beq
b \ = \ \frac{7}{4} \, e^{2\Omega}\, V \left( 1 \ + \ 20 \, \frac{\Omega'}{\phi'}\right) \ .
\eeq

Making use of the explicit solutions in eqs.~\eqref{9D_heterotic_solution} and \eqref{omega_heterotic}, one thus finds
\beq
b \ = \ \frac{8}{3} \, e^{2\Omega}\, V \, \frac{1 \ - \ \frac{1}{2}\ \tanh^2\left( \sqrt{\alpha_H} y\right)}{1 \ + \ 4\ \tanh^2\left( \sqrt{\alpha_H} y\right)} \ ,
\eeq
which is again non negative, so that \emph{the heterotic vacuum is also a perturbatively stable solution of the Einstein--dilaton system} for all allowed choices of boundary conditions at the ends of the interval. The presence of regions where world--sheet or string loop corrections are large, however, makes the lessons of these results less evident for String Theory.

\subsection{Tensor and Vector Perturbations} \label{tensor_9d}

Tensor perturbations are simpler to study, and to this end one only allows a transverse traceless $h_{\mu\nu}$. After a Fourier transform one is thus led to
\beq
h_{\mu\nu}'' \ + \ 8 \, \Omega'\, h_{\mu\nu}'\ + \ m^2\,h_{\mu\nu} \ = \ 0 \ ,
\eeq
which defines a Schr\"odinger problem along the lines of eq.~\eqref{eq_schro}, with $b=0$ and $a=8\Omega'$. Hence, with Dirichlet or Neumann boundary conditions the argument of Section \ref{sec:scalar9d} applies, and one gets a discrete spectrum of masses. Moreover, one can verify that there is a normalizable zero mode with $h_{\mu\nu}$ independent of $z$, which signals that at low energies gravity is effectively nine dimensional.

As in previous cases, vector perturbations entail some mixings, because in this case they originate from transverse $h_{\mu 9}$ and from the traceless combination
\beq
h_{\mu\nu} \ = \ \partial_\mu\, \Lambda_\nu \ + \ \partial_\nu\, \Lambda_\mu \ ,
\eeq
so that
\beq
\partial^\mu\, \Lambda_\mu \ = \ 0 \ .
\eeq
The relevant vector is
\beq
C_\mu \ = \ h_{\mu9} \ - \ \Lambda_\mu' \ ,
\eeq
which satisfies the two equations
\bea
&& \left( p_\mu\,C_\nu \ + \ p_\nu\, C_\mu \right)'\ + \ 8\, \Omega'\, \left( p_\mu\,C_\nu \ + \ p_\nu\, C_\mu \right) \ = \ 0 \ , \nonumber \\
&& m^2 \, C_\mu \ = \ 0  \ ,
\eea
the first of which is clearly solved by
\beq
C_\mu \ = \ C_\mu^{(0)} \ e^{\,-\, 8\, \Omega} \ ,
\eeq
with a constant $C_\mu^{(0)}$. In analogy with the preceding discussion, one might be tempted to identify a massless vector. However, one can verify that, contrary to the case of tensors, this is not associated to a normalizable zero mode. The result is consistent with standard expectations from Kaluza--Klein theory, since the internal manifold has no translational isometry.

\section{The Linear--Dilaton Vacua} \label{sec:subcritical}
Variants of this setting encompass the case of non--critical strings, lower--dimensional compactifications of the ten--dimensional models that we are analyzing in the presence of an internal breathing mode, and also the very important case of cosmological perturbations in four dimensions. In particular, one can describe non--critical strings starting from
\beq
{\cal S}  \ = \  \frac{1}{2\,k_{s}^2}\int d^{d}x\, \sqrt{-g}\left[-\ R\ - \ \frac{4}{d-2}\ (\partial\phi)^2\ - \  V(\phi)  \,
\right] \ , \label{lagrangian_bsb_D}
\eeq
where
\beq
V \ = \ T\, e^{\widetilde{\gamma} \, \phi} \ ,
\eeq
with $\alpha> 0$ for $d > 10$ and $\alpha<0$ for $d < 10$. Moreover
\beq
\widetilde{\gamma} \ = \ \frac{4}{d-2}
\eeq
for the potentials arising from the Polyakov measure~\cite{polyakov}.

The equations of motion read
\bea
\Box\phi  &-& \frac{d-2}{8}\, \frac{\partial V}{\partial \phi} \ = \ 0 \ , \nonumber \\
R_{MN} &+& \frac{4}{d-2} \, \pr_M\phi\, \pr_N\phi\ + \ \frac{1}{d-2}\ g_{MN}\, V(\phi) \ = \ 0 \ ,
\eea
and the corresponding perturbed fields take again the form
\begin{eqnarray}
&& ds^2 \ = \ e^{2\,\Omega(z)} \Big[\eta_{MN}\,+\,h_{MN}(x,z) \Big] dx^M\,dx^N  \ , \nonumber \\
&& \phi \ = \ \phi(z) \ + \ \varphi(x,z)
\ , \label{9D_solution_conformal_d}
\end{eqnarray}
while the perturbed Ricci curvature can be extracted from
\bea
R_{MN} &=& (d-2)\, \nabla_M\, \nabla_N\, \Omega \ + \ (\eta_{MN}+h_{MN})\, \nabla^A\,\nabla_A\,\Omega \nonumber \\ &-& (d-2)\ \Big( \nabla_M\, \Omega\,\nabla_N\, \Omega \ - \ (\eta_{MN}+h_{MN}) \,\nabla^A\,\Omega \,\nabla_A\,\Omega \Big) \nonumber \\
&+& \frac{1}{2} \, \Big[(\Box\,+\, \partial_z^2)\, h_{MN} \ -\ \nabla_M \left(\nabla \cdot h\right)_{N}\ - \ \nabla_N \left(\nabla \cdot h\right)_{M}\ +\ \nabla_M\,\nabla_N \, h_L{}^L \Big] \ .
\eea
The unperturbed equations of motion then read
\bea
\Omega''\ +\ (d-2)\,(\Omega')^2\ +\ \frac{e^{2\Omega}}{d-2}\,V&=&0\ ,\nonumber \\
(d-1)\,\Omega''\ +\ \frac{e^{2\Omega}}{d-2}\,V \ +\ \frac{4}{d-2}\,(\phi')^2&=&0\ , \nonumber \\
\phi''\ +\ (d-2)\,\Omega'\phi'\ - \ \frac{(d-2)}{8}\ e^{2\Omega}\, V_\phi(\phi) &=&0\ , \label{eqfi_d}
\eea
where, as in the preceding section, $V$ and $V_\phi$ denote the potential and its derivatives, computed on the classical vacuum. The background solution that we are after in this section is the linear--dilaton vacuum discussed in~\cite{lindil_1}, which is an exact classical solution of String Theory, to all orders in $\alpha'$. The corresponding field profiles are
\beq
\Omega \ = \  a \, z \ , \qquad
\phi   \ = \ - \ \frac{a}{2} \, (d-2)\,z \ , \label{lindilsol_subc}
\eeq
where the parameter $a$ is determined by the potential in eq.~\eqref{lagrangian_bsb_D}
according to
\beq
e^{2\Omega}\,V_0 \ = \ - \ (d-2)^2\, a^2 \ . \label{lindilpot_subc}
\eeq

Proceeding as in the preceding section, one is thus led to the equation of scalar perturbations
\bea
A'' &+& A'\left[3(d-2)\, \Omega'\ - \ \frac{(d-2)}{4\,\phi'} \ e^{2\Omega}\, V_{\phi} \right] \nonumber \\ &+&
A\left[m^2 \ - \ \frac{2(d-3)}{(d-2)} \ e^{2\Omega}\, V \ - \ \frac{(d-2)(d-3)}{4} \, e^{2\Omega}\, \Omega'\, \frac{V_\phi}{\phi'} \right] \ = \ 0 \ , \label{second_order_9d_d}
\eea
where the relations between $C$, $A$ and $\varphi$ are now
\bea
C &=& - \ (d-3)\,A \ , \nonumber \\
\varphi &=& - \ \frac{(d-2)^2}{8\,\phi'} \left[A'\ + \ (d-3)\,A\,\Omega'\right] \ . \label{phi_ACphi_d}
\eea
In a similar fashion, tensor perturbations are described by
\beq
h_{\mu\nu}'' + \ (d-2) \Omega'\,h_{\mu\nu}' \ + \ m^2\,h_{\mu\nu} \ = \ 0 \ . \label{d_tensor}
\eeq
For the linear dilaton solution of eqs.~\eqref{lindilsol_subc} and \eqref{lindilpot_subc}, eqs.~\eqref{second_order_9d_d} and \eqref{d_tensor} become identical and take the simple form
\beq
A'' \ + \ A'\,(d-2)\,a \ + \ m^2\, A \ = \ 0 \ .
\eeq
Retracing the preceding steps one is thus led to define
\beq
A(z) \ = \ \Psi(z) \, e^{\,- \,\frac{d-2}{2}\,a\, z} \ ,
\eeq
and then
\beq
\frac{d^2 \Psi}{dz^2} \ + \ \left[m^2 \ - \ \frac{a^2(d-2)^2}{4} \right] \Psi \ = \ 0 \ , \label{schroed_subc}
\eeq
which admits bounded solutions only if
\beq
m^2\ > \ \frac{(d-2)^2\,a^2}{4} \ .
\eeq
Therefore, this linear--dilaton background is stable. However, it has several problems. To begin with, although eqs.~\eqref{lindilsol_subc} and \eqref{lindilpot_subc} correspond to an exact solution of tree--level String Theory to all orders in $\alpha'$, there is a region of strong coupling for large negative values of $z$, and moreover the preceding argument shows that there are no massless scalar or tensor modes around this vacuum. In addition, from a lower--dimensional perspective there is a continuum of massive excitations.

\section{The Dudas--Mourad Cosmological Solutions in Ten Dimensions}\label{sec:climbing}

Our starting point is now the analytic continuation under $y \to i t$, and consequently under $z \to i \eta$, of the classical solutions of Section \ref{sec:dm_vacua}. For the ten--dimensional Sugimoto orientifold~\cite{sugimoto} one thus obtains, in the Einstein frame,
\begin{eqnarray}
\phi &=& - \ \frac{3}{4}\ \alpha_O t^2\ + \ \frac{1}{3}\ \ln \left|{\alpha_O} t^2\right| \ + \ \Phi_0 \ , \nonumber \\
ds_O^2 &=& \left|{\alpha_O} \,t^2\right|^\frac{1}{18} e^{\alpha_O \frac{t^2}{8}}\  d{\mathbf{x}} \cdot d{\mathbf{x}}
\ - \ e^{-\frac{3}{2}\, \Phi_0}\,\left|{\alpha_O} t^2\right|^{-\frac{1}{2}} e^{\frac{9}{8}\, \alpha_O t^2} dt^2 \  \label{s19}\ , \label{9D_orientifold_solution_cosmo}
\end{eqnarray}
where the parametric time $t$ takes values, as usual for a decelerating Cosmology with an initial singularity, in $(0,\infty)$. The corresponding Einstein--frame solution for the $SO(16) \times SO(16)$ {heterotic model} of~\cite{so16xso16} reads
\begin{eqnarray}
\phi &=& \frac{1}{2}\ \ln \sin\left|\sqrt{\alpha_H} t \right| \ + \ 2\, \ln \cos\left|\sqrt{\alpha_H} t\right| \ + \ \Phi_0 \ , \nonumber \\
ds_H^2 &=& \left|\sin\left(\sqrt{\alpha_H} t \right)\right|^\frac{1}{12}\,\left|\cos\left(\sqrt{\alpha_H} t \right)\right|^{-\frac{1}{3}} \  d{\mathbf{x}} \cdot d{\mathbf{x}}
\nonumber \\ &-& e^{-\frac{5}{2}\, \Phi_0}\,\left|\sin\left(\sqrt{\alpha_H} t \right)\right|^{-\frac{5}{4}}\,\left|\cos\left(\sqrt{\alpha_H} t \right)\right|^{-5} \,
 dt^2 \  \label{s1}\ , \label{9D_heterotic_solution_cosmo}
\end{eqnarray}
where $0 < \sqrt{\alpha_H} t < \frac{\pi}{2}$.
Both cosmologies have a nine--dimensional Euclidean symmetry, and in both cases, as shown in~\cite{climbing}, the dilaton is forced to emerge from the initial singularity climbing up the potential. In this fashion it reaches an upper bound, which translates into an upper limit on the string coupling, before it begins its descent.

It is convenient to recast these expressions in conformal time according to
\begin{eqnarray}
&& ds^2 \ = \ e^{2\,\Omega(\eta)} \left( d{\mathbf{x}} \cdot d{\mathbf{x}} \ - \ d\eta^2 \right) \ , \nonumber \\
&& \phi \ = \ \phi(\eta)
\ . \label{9D_conformal_cosmo}
\end{eqnarray}
In the orientifold case the conformal time $\eta$ is obtained integrating the relation
\beq
d \eta \ = \ \left| \sqrt{\alpha_O}\, t \right|^{-\,\frac{5}{9}}\, e^{\,-\,\frac{3}{4}\, \Phi_0} \, e^{\frac{\alpha_O}{2}\, t^2}\ d t \ ,
\eeq
and
\beq
e^{2\,\Omega(\eta)} \ = \ |\sqrt{\alpha_O} \,t|^\frac{1}{9} e^{\frac{\alpha_O}{8} t^2} \ , \label{omega_orientifold_cosmo}
\eeq
while in the heterotic case
\beq
d\eta \ = \ \left|\sin\left(\sqrt{\alpha_H} t \right)\right|^{-\frac{2}{3}}\,\left|\cos\left(\sqrt{\alpha_H} t \right)\right|^{-\frac{7}{3}}\, e^{\,-\,\frac{5}{4}\, \Phi_0} \ d t\ ,
\eeq
and
\beq
e^{2\,\Omega(\eta)} \ = \ \left|\sin\left(\sqrt{\alpha_H} t \right)\right|^\frac{1}{12}\,\left|\cos\left(\sqrt{\alpha_H} t \right)\right|^{-\frac{1}{3}} \ . \label{omega_heterotic_cosmo}
\eeq
In both models one can choose the range of $\eta$ to be $(0,\infty)$, with the initial singularity at the origin. Moreover, as we have stressed, in both cases the dilaton, and thus the string coupling, is bounded from above, so that string loops are in principle under control. However, there is still a singularity at $t=0$ where string world--sheet corrections, which are not taken into account in this Supergravity approximation, are expected to play an important role~\cite{cond_dud}.

\subsection{Tensor Perturbations} \label{sec:tensor_climbing}

The issue at stake, here and in the following sections, is whether or not solutions determined by arbitrary initial conditions provided some time after the initial singularity can grow in the future evolution of the Universe.
This can be done relatively simply at large times, which translate into large values of the conformal time $\eta$, where many expressions simplify. Moreover, for finite values of $\eta$ the geometry is regular and the coefficients in eq.~\eqref{cosmo_tensor} are bounded, so that the solutions are also not singular. However, an ${\cal O}(1)$ growth is relevant for perturbations, and therefore we shall begin with the asymptotics and then, at the end of the section, we shall also take a global look at the problem.

In the ten--dimensional orientifold and heterotic models of interest, using spatial Fourier transforms and proceeding as in the preceding section, one can show that tensor perturbations evolve according to
\beq
h_{ij}'' \ + \ 8 \, \Omega'\, h_{ij}'\ + \ {\bf k}^2\,h_{ij} \ = \ 0 \ , \label{cosmo_tensor}
\eeq
where ``primes'' denote derivatives with respect to the conformal time $\eta$.
Let us begin by noting that, for all exponential potentials
\beq
V \ = \ T\, e^{\gamma_E\,\phi}  \label{Vgammae}
\eeq
with $\gamma_E \geq \frac{3}{2}$, and therefore for the orientifold potentials of~\cite{bsb} and~\cite{u32}, which have $\gamma_E=\frac{3}{2}$ and are ``critical'' in the sense of~\cite{climbing}, but also for the $SO(16) \times SO(16)$ heterotic string of~\cite{so16xso16}, which has $\gamma_E=\frac{5}{2}$ and is ``super--critical'' in the sense of~\cite{climbing}, the solutions of the background equations
\bea
\Omega''+8(\Omega')^2\ - \ \frac{e^{2\Omega}}{8}\,V&=&0\ ,\nonumber\\
9\Omega''\ - \ \frac{e^{2\Omega}}{8}\,V \ +\ \frac{1}{2}\,(\phi')^2&=&0\ ,\label{eqfi_cosmo}\\
\phi''\ +\ 8\Omega'\phi'\ + \ \gamma_E\, e^{2\Omega}\,V &=&0 \nonumber
\eea
are dominated, for large values of $\eta$, by
\beq
\phi \ \sim \ - \  \frac{3}{2}\ \log( \sqrt{\alpha_H}\,\eta) \ , \qquad \Omega \  \sim \  \frac{1}{8}\ \log ( \sqrt{\alpha_H}\,\eta) \ , \qquad e^{2\Omega}\, V \ \sim \ 0 \ . \label{largetacritical}
\eeq
In the picture of~\cite{climbing}, in this region the scalar field has overcome the turning point and is descending the potential, so that the supergravity approximation is expected to be reliable, but the potential contribution is manifestly negligible only for the ``super--critical'' heterotic potential~\eqref{pot52}, where $e^{2\Omega}\,V$ decays faster than $\frac{1}{\eta^2}$ for large $\eta$. However, the result also applies for $\gamma_E=\frac{3}{2}$, which marks the onset of the ``climbing behavior''. This can be appreciated retaining subleading terms, which results into
\beq
\phi \sim - \frac{3}{2} \, \log \left( \sqrt{{\alpha_{O}}}\eta \right) - \frac{5}{6} \, \log\left(\log\left( \sqrt{{\alpha_O}}\eta\right)\right) \ , \quad \Omega \sim \frac{1}{8} \, \log \left( \sqrt{{\alpha_{O}}}\eta\right) + \frac{1}{8} \, \log\left(\log\left( \sqrt{{\alpha_{O}}}\eta\right)\right)\ , \label{trickyphiomega}
\eeq
so that the potential decays as
\beq
e^{2\,\Omega}\, V \ \sim \ \frac{T}{\left( \sqrt{{\alpha_O}}\eta \right)^{2}\,[2\,\log( \sqrt{{\alpha_O}}\eta)]} \ , \label{trickyV}
\eeq
which is faster than $\frac{1}{\eta^2}$.

Notice that a similar behavior, but with the scalar climbing up the potential,
\beq
\phi \ \sim \  \frac{3}{2}\ \log( \sqrt{\alpha_{O,H}}\,\eta) \ , \quad \Omega \  \sim \  \frac{1}{8}\ \log ( \sqrt{\alpha_{O,H}}\,\eta) \ , \quad e^{2\Omega}\, V \ \sim \ 0 \ , \label{smalletacritical}
\eeq
also emerges for small values of $\eta$, for all $\gamma_E \geq \frac{3}{2}$, and thus in all orientifold and heterotic models of interest.
However, these expressions are less compelling, since they concern the onset of the climbing phase. The potential is manifestly subdominant for small values of $\eta$, but curvature corrections, which would be important in this region, are not simply taken into account.

In conclusion, for $\gamma_E \geq \frac{3}{2}$ and for large values of $\eta$ eq.~\eqref{largetacritical} holds and eq.~\eqref{cosmo_tensor}, which describes tensor perturbations, therefore approaches
\beq
h_{ij}'' \ + \ \frac{1}{\eta} \, h_{ij}'\ + \ {\bf k}^2\,h_{ij} \ = \ 0 \ . \label{tensor_cosmo_10}
\eeq
Consequently, for $\mathbf{k} \neq 0$
\beq
h_{ij} \ \sim \ A_{ij}\, J_0(k\eta) \ + \ B_{ij}\, Y_0(k \eta) \ .\,
\eeq
and the oscillations are damped for large times, so that \emph{no instabilities arise}.
\begin{figure}[ht]
\begin{center}
\includegraphics[width=90mm]{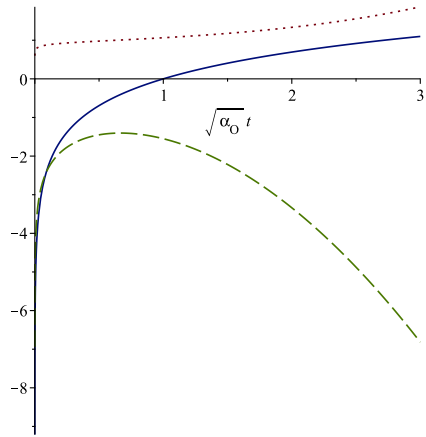}
\vspace*{-5.5truecm}
\end{center}
\caption{\small The scale factor $e^\Omega$ (red, dotted), the unstable homogeneous tensor mode (blue) and the scalar field $\phi$ (green, dashed) as a function of $\sqrt{\alpha_o}t$, where the parametric time $t$ is defined in Section~\ref{sec:climbing}.}
\label{fig:instability_tensor}
\end{figure}
On the other hand, an intriguing behavior emerges for ${\mathbf{k}}  = 0$, which is worthy of some attention in our opinion. In this case the solution of eq.~\eqref{tensor_cosmo_10} implies that
\beq
h_{ij} \ \sim \ A_{ij} \ + \ {B_{ij}} \ \log\left(\frac{\eta}{\eta_0}\right)  \ , \label{logk0}
\eeq
and therefore \emph{spatially homogeneous tensor perturbations experience in general a logarithmic growth}. This result indicates that \emph{homogeneity is preserved while isotropy is generally violated} in the ten--dimensional ``climbing--scalar'' cosmologies~\cite{climbing} that emerge, in String Theory, in the presence of supersymmetry breaking.

One can actually get a global picture of the phenomenon: eq.~\eqref{tensor_cosmo_10} can be simply  solved in terms of the parametric times of~\cite{dm_9Dsolution} for ${\mathbf{k}}=0$, and
\beq
h_{ij} \ = \ A_{ij} \ + \ B_{ij}\, \log\left(\sqrt{\alpha_O} t \right) \ , \label{inst_hom_orient}
\eeq
for the $USp(32)$ and $U(32)$ orientifold models, while (in this case $\ 0\ < \ \sqrt{\alpha_H} t \ < \ \frac{\pi}{2}$)
\beq
h_{ij} \ = \ A_{ij} \ + \ B_{ij}\, \log\left[\tan\left(\sqrt{\alpha_H} t\right) \right] \ , \label{inst_hom_heter}
\eeq
for the $SO(16) \times SO(16)$ heterotic model. These results are qualitatively similar, if one takes into account the limited range of $t$ in the heterotic potential~\eqref{pot52}, and typical behaviors related to eq.~\eqref{inst_hom_orient} are displayed in fig.~\ref{fig:instability_tensor}. The general lesson is that perturbations acquire quickly ${\cal O}(1)$ variations toward the end of the climbing phase, where curvature corrections to Supergravity do not dominate the scene anymore, providing some additional support to the present analysis.
\emph{We are therefore inclined to speculate that the violation of isotropy that we are exhibiting could have left behind some space--time dimensions at a very Early Epoch, opening the way to an effective four--dimensional Universe.}~\footnote{This result is implied by the ``(super)critical'' potentials of eqs.~\eqref{pot32} and \eqref{pot52}, and resonates at least with some previous investigations~\cite{anagnostopoulos} of matrix models for the type--IIB theory~\cite{eguchi}.} While perturbation theory is at most a clue to this effect, the resulting picture is clearly enticing, and moreover, as we shall see shortly, the dynamics becomes potentially richer and stable in lower dimensions, where other branes that become space filling can inject an inflationary phase devoid of this type of instability.

These amusing properties apply to all ``(super)critical''
exponential potentials~\footnote{Notice that the exponent $\gamma$ in~\cite{climbing} reflects non--conventional normalizations that set the transition to this type of behavior at $\gamma=1$ for all dimensions $d$.}
\beq
V \ = \ T \, e^{\gamma_E\,\phi}  \label{pot_gammae}
\eeq
with $\gamma_E \geq \frac{3}{2}$, where the climbing phenomenon occurs. In all these cases, the potential energy becomes subdominant in the large--$\eta$ descending phases, so that some aspects of these systems are captured by fluids with an equation of state
\beq
p \ = \ \gamma\, \rho \label{eq_of_state} \ ,
\eeq
with a parameter $\gamma$ that approaches one from below. This limiting behavior emerges whenever a positive potential becomes subdominant, since in general for a scalar field
\beq
\frac{p}{\rho} \ = \ \frac{T \ - \ V}{T \ +\ V} \ ,
\eeq
with $T$ and $V$ the corresponding kinetic and potential energies. It is thus instructive to take a cursory look at corresponding Freedman--Robertson--Walker cosmologies in $d$ dimensions, for which
\beq
ds^2 \ = \ -\ dt^2 \ +  \left(\frac{t}{t_0}\right)^\frac{4}{(d-1)(1+\gamma)} \ d {\mathbf x} \cdot d {\mathbf x} \ ,
\eeq
so that in conformal time
\beq
ds^2 \ = \ \left(\frac{\eta}{\eta_0}\right)^\frac{4}{(d-1)(1+\gamma) \ - \ 2} \Big[ -\ d\eta^2 \ +  \ d {\mathbf x} \cdot d {\mathbf x} \Big] \ .
\eeq
Tensor perturbations would evolve in this background according to eq.~\eqref{tensor_d_cosmo}, a slight modification of eq.~\eqref{cosmo_tensor}, as we shall see in Section~\ref{sec:general_exponential}, and the result for homogeneous perturbations would be
\beq
h_{ij} = A_{ij}  \ + \ B_{ij} \ \eta^{\frac{(d-1)(\gamma-1)}{(d-1)(\gamma+1)-2}} \ .
\eeq
Therefore, in the standard region $\gamma<1$ one would always end up with bounded perturbations at large times, even for ${\mathbf k} = 0$. On the other hand, \emph{the behavior in eq.~\eqref{logk0} would always present itself for $\gamma=1$}, while negative potentials would give rise to $\gamma>1$, and thus to worse instabilities.

To reiterate, the instability in the homogeneous tensor modes that we have exhibited for climbing scalars with $\gamma_E \geq \frac{3}{2}$ signals a problem with isotropy that is common to all ten--dimensional string models with broken supersymmetry. As we have already stressed, this points naturally to an awaited tendency toward lower--dimensional space times, albeit without a selection criterion for the resulting dimension $d < 10$.  Hence, if the picture that we are advocating were correct, the selection of a four--dimensional Universe in String Theory would be an accident, no more and no less than in other contexts, but with hints from the breaking of Supersymmetry that point to its inevitable emergence. Before proceeding to examine lower--dimensional settings, let us take a look at scalar perturbations in ten dimensions.

\subsection{Scalar Perturbations} \label{sec:scalar_climbing}

Scalar perturbations exhibit a very different behavior in the presence of the exponential potentials~\eqref{pot_gammae} with $\gamma_E \geq \frac{3}{2}$, and in particular of those in eqs.~\eqref{pot32} and \eqref{pot52} that reflect the breaking of Supersymmetry in ten--dimensional String Theory.
Our starting point is the analytic continuation of eq.~\eqref{second_order_9d} under $z \to i \,\eta$, which reads
\beq
A'' +\left(24\, \Omega'\ + \ \frac{2}{\phi'} \ e^{2\Omega}\, V_{\phi} \right)A' \ + \
\left({\mathbf k}^2 \ + \ \frac{7}{4} \ e^{2\Omega}\, V \ + \ 14 \, e^{2\Omega}\, \Omega'\, \frac{V_\phi}{\phi'} \right)A \ = \ 0 \ . \label{second_order_9d_scalar_cosmo}
\eeq
As in eq.~\eqref{cosmo_tensor}, $m^2$ was also replaced with $-{\mathbf k}^2$, which originates from a spatial Fourier transform, and ``primes'' denote again derivatives with respect to the conformal time $\eta$.

As we have stressed in the preceding section, the potential is subdominant in eq.~\eqref{second_order_9d_scalar_cosmo} for $\gamma_E \geq \frac{3}{2}$, which leads to the asymptotic behaviors of eqs.~\eqref{smalletacritical} during the climbing phase, and of eqs.~\eqref{largetacritical} during the descending phase. As a result, during the latter eq.~\eqref{second_order_9d_scalar_cosmo} reduces to
\beq
A''\  +\ \frac{3}{\eta}\, A' \ + \ k^2\, A \ = \ 0 \ ,
\eeq
whose solution is
\beq
A \ = \ A_1 \ \frac{J_1(k\,\eta)}{\eta} \ + \ A_2 \ \frac{Y_1(k\,\eta)}{\eta} \ .
\eeq
For ${\bf k} \neq 0$ the amplitude always decays proportionally to $(\eta)^{-\frac{3}{2}}$, while for ${\bf k}=0$ the two independent solutions of eq.~\eqref{second_order_9d_scalar_cosmo} are dominated by
\beq
A \ = \ A_3 \ + \ A_4 \ \frac{1}{\eta^2} \ .
\eeq
Therefore, \emph{scalar perturbations do not grow in time}, even for the homogeneous mode with ${\mathbf{k}}=0$, for $\gamma_E \geq \frac{3}{2}$, and thus, in particular, for the $USp(32)$ and $U(32)$ orientifold models of~\cite{bsb} and \cite{u32}, and also for the $SO(16) \times SO(16)$ heterotic model of~\cite{so16xso16}.

\section{A Family of Cosmologies in $d$ Dimensions} \label{sec:general_exponential}

In Section~\ref{sec:tensor_climbing} we have noticed that homogeneous tensor perturbations develop an instability in the ten--dimensional potentials of eqs.~\eqref{pot32} and \eqref{pot52}, and we have emphasized that this breakdown of isotropy could have ignited the transition to a four--dimensional Universe. Here we would like to explore cosmological perturbations in the richer dynamical settings that become available for $d < 10$. The picture that we have in mind is that these more general potentials could have driven the scalar field along them during its descent, and in this spirit we shall explore whether instabilities arise in this dynamical regime.

We have actually in mind two types of action principles. The first is the naive extension to $d<10$ of eq.~\eqref{lagrangian_bsb},
\beq
{\cal S}  \ = \  \frac{1}{2\,k_{d}^2}\int d^{d}x\, \sqrt{-g}\left[-\ R\ - \frac{4}{d-2}\ (\partial\phi)^2\ - \  T\,e^{\widetilde{\gamma}\,\phi} \,
\right] \ , \label{lagrangian_cosmo_D}
\eeq
considered however for a ``mild'' (in a sense to be specified shortly) positive exponential potential characterized by a parameter $\widetilde{\gamma}$. In contrast, the ten--dimensional models contained two types of ``hard'' potentials, which we first met in eqs.~\eqref{pot32} or \eqref{pot52}. We shall leave these contributions implicit here: their main lesson, the inevitable presence of an early climbing phase, was already taken into account. The second type of action,
\beq
{\cal S}  \ = \  \frac{1}{2\,k_{d}^2}\int d^{d}x\, \sqrt{-g}\left[-\ R\ - \frac{1}{2}\ (\partial\chi)^2\ - \  T \,e^{\gamma_d\,\chi} \,
\right] \ , \label{lagrangian_breathing_D}
\eeq
captures lower--dimensional mixings between the dilaton and the breathing mode of the internal manifold~\cite{integrable}. This second option is relevant after the compactification, up to another orthogonal mixing of dilaton and breathing mode that, as in~\cite{integrable}, we shall assume to be somehow stabilized and foreign to the dynamics. Clearly, we cannot nearly claim full control of all this interesting dynamics, not even within Supergravity, but we shall attempt nonetheless to extract potentially useful indications from it, within this assumption.

Notice also that the two actions of eqs.~\eqref{lagrangian_cosmo_D} and \eqref{lagrangian_breathing_D} can be simply turned into one another by a redefinition of $\phi$, which also connects $\gamma_d$ to $\widetilde{\gamma}$ according to~\footnote{This redefinition, determined by eq.~\eqref{lagrangian_breathing_D}, refers to a ``string--normalized'' scalar and affects the values of $\gamma_\ell$ and $\gamma^{(c)}$ below. For a canonically normalized scalar one should include another factor $\sqrt{2} \,k_d$ in $\gamma^{(c)}$ and also in $\gamma_{\ell}$ below.}
\beq
\gamma_d \ = \ \widetilde{\gamma} \ \sqrt{\frac{d-2}{8}} \ .
\eeq
One can thus explore their indications simultaneously, starting from the background equations, which in these lower--dimensional settings read
\bea
&&\Omega''\ +\ (d-2)\,(\Omega')^2\ -\ \frac{e^{2\Omega}}{d-2}\,V \ = \ 0\ ,\nonumber\\
&&(d-1)\,\Omega''\ -\ \frac{e^{2\Omega}}{d-2}\,V \ +\ \frac{4}{d-2}\,(\phi')^2 \ = \ 0\ ,\nonumber\\
&&\phi''\ +\ (d-2)\,\Omega'\phi'\ + \ \frac{(d-2)}{8}\ e^{2\Omega}\, V_\phi(\phi)  \ = \ 0\ . \label{cosmo_eqs_d}
\eea
Starting from the results of Section~\ref{sec:subcritical}, one can also show that the tensor perturbations of the Lagrangian~\eqref{lagrangian_cosmo_D} satisfy
\beq
h_{\mu\nu}'' \ +\  (d-2) \Omega'\,h_{\mu\nu}' \ + \ {\mathbf{k}}^2\,h_{\mu\nu} \ = \ 0 \ , \label{tensor_d_cosmo}
\eeq
while the corresponding scalar perturbations satisfy
\bea
A'' &+& A'\left[3(d-2)\, \Omega'\ +\ \frac{(d-2)}{4\,\phi'} \ e^{2\Omega}\, V_{\phi} \right] \nonumber \\ &+&
A\left[{\mathbf{k}}^2 \ + \ \frac{2(d-3)}{(d-2)} \ e^{2\Omega}\, V \ + \ \frac{(d-2)(d-3)}{4} \, e^{2\Omega}\, \Omega'\, \frac{V_\phi}{\phi'} \right] \ = \ 0 \ . \label{scalar_d_cosmo}
\eea

As we have anticipated, we shall focus on dynamics that are dominated by a single ``mild'' exponential term, so that
\beq
V_\phi \ = \ \widetilde{\gamma}\, V \ . \label{pos_mild}
\eeq
Letting
\beq
\gamma_{\ell} \ =\  \frac{4}{d-2} \ , \qquad
\gamma{^{(c)}} \ =\  \frac{4\,\sqrt{d-1}}{d-2} \ , \label{gammac}
\eeq
the asymptotic solutions of eqs.~\eqref{cosmo_eqs_d} for large values of $\eta$, in the region where the scalar descends the potential, are quite different from what we saw in preceding sections, and
\bea
\Omega' &=& \frac{1}{\left(\frac{\widetilde{\gamma}}{\gamma_{\ell}}\right)^2 \ - \ 1 } \ \frac{1}{\eta} \ , \nonumber \\
 \phi'&=& - \ \frac{\widetilde{\gamma}}{8} \ \frac{(d-2)^2}{\left(\frac{\widetilde{\gamma}}{\gamma_{\ell}}\right)^2 \ - \ 1 } \ \frac{1}{\eta} \ , \nonumber \\
e^{2\Omega}\ V &=& \frac{(d-2)^3}{16} \ \frac{{\gamma^{(c)\,2}} \ - \ \widetilde{\gamma}^2}{\left[\left(\frac{\widetilde{\gamma}}{\gamma_{\ell}}\right)^2 \ - \ 1\right]^2}\ \frac{1}{\eta^2} \ . \label{sol_cosmo_d}
\eea
These results describe the Lucchin--Matarrese attractor~\cite{lm} in $d$ dimensions, a special solution that captures the large--time behavior of all solutions of the gravity--scalar system in the presence of ``mild'' positive exponential potentials~\footnote{Exponential potentials in Cosmology have a long history, and some relevant previous work can be found in~\cite{exponential_cosmology}.} This qualification, as we can now see, translates into the condition
\beq
\widetilde{\gamma} \ < \ \gamma^{(c)} \ , \label{criticality}
\eeq
since the solutions in eq.~\eqref{sol_cosmo_d} exist, for positive potentials, only within this range.

Making use of these results in eqs.~\eqref{tensor_d_cosmo} and \eqref{scalar_d_cosmo} leads to two very similar equations,
\bea
h_{\mu\nu}'' &+&  \frac{a_T}{\eta} \ h_{\mu\nu}' \ + \ {\mathbf{k}}^2\,h_{\mu\nu} \ = \ 0 \ . \label{tensor_d_cosmo_exp} \\
A'' &+& \frac{a_S}{\eta}\, A' \ + \
{\mathbf{k}}^2\, A  \ = \ 0 \ , \label{scalar_d_cosmo_exp}
\eea
where
\bea
a_T &=&   \frac{(d-2) }{\left(\frac{\widetilde{\gamma}}{\gamma_\ell}\right)^2 \ - \ 1 } \ ,  \\
a_S &=&   \frac{(d-4) \ + \ 2\left(\frac{\widetilde{\gamma}}{\gamma_\ell}\right)^2}{\left(\frac{\widetilde{\gamma}}{\gamma_\ell}\right)^2 \ - \ 1} \ = \ 2 \ + \ a_T \ .
\eea
Denoting the corresponding fields collectively as $Y$, the solutions can be presented in the form
\beq
Y(\eta) \ = \ C_1 \ \left|\eta\right|^\frac{1-a}{2} J_{\left| \frac{1-a}{2} \right|}\left(k\,|\eta|\right) \ + \ C_2 \ \left|\eta\right|^\frac{1-a}{2} Y_{\left| \frac{1-a}{2} \right|}\left(k\,|\eta|\right)
\eeq
for ${\mathbf{k}}\neq 0$, and
\beq
Y(\eta) \ = \ D_1 \, \left|\eta\right|^{1-a} \ + \ D_2
\eeq
for ${\mathbf{k}}=0$,
with $a=a_T$ for tensor perturbations and $a=a_S$ for scalar ones. Notice that for $\widetilde{\gamma}> \gamma_\ell$ the Universe undergoes a decelerated expansion, and therefore we are interested in the behavior as $\eta \to \infty$, while if $\widetilde{\gamma}<\gamma_\ell$ it undergoes an accelerated expansion, and therefore we are interested in the behavior as $\eta \to 0$.

Leaving aside momentarily the case $\widetilde{\gamma}=\gamma_\ell$, to which we shall return in the next section, one can simply verify that perturbations never grow (when $d>2$) if eq.~\eqref{criticality}, which is instrumental for the existence of the Lucchin--Matarrese attractor with positive potentials, holds. To reiterate, \emph{no instabilities arise, even for homogeneous perturbations, in the presence of a ``mild'' exponential potential.}
Notice also that, in the picture subsumed by eq.~\eqref{lagrangian_breathing_D}, $\gamma^{(c)}$ becomes the ``critical'' value of~\cite{integrable}.

As in Section~\ref{sec:tensor_climbing}, one can get a simple global view of the homogenous solutions of eq.~\eqref{tensor_d_cosmo}. This is determined by the first integral of eqs.~\eqref{tensor_d_cosmo_exp} and \eqref{scalar_d_cosmo_exp}
\beq
d h_{ij} \ = \ C_{ij}\, e^{-(d-2)\Omega}\, d\eta \ ,
\eeq
where $C_{ij}$ is a constant matrix, an equation that
can be simply integrated working in terms of the parametric time of~\cite{dm_9Dsolution}.
The starting point in this case is provided, in the notation of~\cite{climbing}, by
\bea
e^{2\Omega} &=& \left|\sinh\left(\frac{\tau}{2}\,\sqrt{1-\gamma^2}\right) \right|^\frac{2}{(1+\gamma)(d-1)} \ \left| \cosh\left(\frac{\tau}{2}\,\sqrt{1-\gamma^2}\right) \right|^\frac{2}{(1-\gamma)(d-1)} \\
d\eta&=& \frac{d\tau}{M}\, \sqrt{\frac{d-2}{d-1}} \ e^{-2\gamma\varphi_0}\left|\sinh\left(\frac{\tau}{2}\,\sqrt{1-\gamma^2}\right) \right|^{- \frac{\gamma(d-1)+1}{(1+\gamma)(d-1)}} \ \left| \cosh\left(\frac{\tau}{2}\,\sqrt{1-\gamma^2}\right) \right|^\frac{\gamma(d-1)-1}{(1-\gamma)(d-1)} \ , \nonumber
\eea
but the scalar field and the parametric time were redefined there in such a way that the transition to the ``climbing behavior'' occurred, for any value of the dimension $d$, for $\gamma=1$. The translation into the present notation rests on the redefinitions
\beq
\tau \ \longrightarrow \ t\, \sqrt{\alpha_{O,H}} \ \sqrt{\frac{2(d-1)}{d-2}} \ , \qquad M \ \longrightarrow \ \alpha_{O,H} \, \sqrt{2} \ , \qquad \gamma \ \longrightarrow \ \frac{\widetilde{\gamma}}{\gamma^{(c)}} \ , \label{redefinitions}
\eeq
and taking them into account leads, for homogeneous tensor perturbations, to the result
\beq
h_{ij} \ = \ A_{ij} \ + \ {B_{ij}} \ \log\left[\tanh\left(\sqrt{\alpha_{O,H}} \, t\,\sqrt{\frac{d-1}{2(d-2)}}\sqrt{1\ - \ \left(\frac{\widetilde{\gamma}}{\gamma^{(c)}}\right)^2}\right)\right] \ ,
\eeq
where $\alpha_{O,H}$ stands for $\alpha_O$ in the orientifold case and for $\alpha_H$ in the heterotic case.
This expression displays clearly that \emph{during the descent, the region of interest for the present considerations, this homogeneous perturbation does not grow appreciably, so that no instabilities arise in homogenous tensor perturbations for $\widetilde{\gamma} < \gamma^{(c)}$.}
This is as expected, due to the attractor nature of the Lucchin--Matarrese solution.

What are the relevant values of $\frac{{\gamma}_d}{\gamma^{(c)}}$ that result from lower--dimensional branes? This was discussed in~\cite{integrable}, under the assumptions that were spelled out after eq.~\eqref{lagrangian_breathing_D}. In our current notation the answer in four dimensions is
\beq
\frac{{\gamma}_d}{\gamma^{(c)}} \ = \ \frac{1}{12} \left( p \ + \ 9 \ - \ 6\, \sigma \right) \ , \label{gammaovergammac}
\eeq
for a $p$-brane that couples to the dilaton proportionally to $e^{-\sigma \phi}$. Lower--dimensional non--BPS branes in the string models of interest were discussed in~\cite{dms}, and for instance in four dimensions there is a non--BPS $D3$ brane of this type, with $\frac{{\gamma}_d}{\gamma^{(c)}}=\frac{1}{2}$. In general, there is a whole zoo of branes in $d<10$, which were studied in detail by Bergshoeff, Riccioni and others~\cite{berg_ric}, so that there are in principle many options for this type of stable dynamics.

Leaving aside the need for further contributions to the potential that are needed to comply with well--known bounds on the tensor--to--scalar ratio $r$~\cite{PLANCK}, when combining in four dimensions a ``hard'' exponential with a ``mild'' term one stumbles on some amusing numerology. The values of $\widetilde{\gamma}$ translate indeed into the spectral index for primordial scalar perturbations, according to
\beq
n_S \ = \ \frac{\gamma^{(c)\,2} \ - \ 9\,{\gamma_d}^2}{\gamma^{(c)\,2} \ - \ 3\,{\gamma_d}^2} \ ,
\eeq
and $\frac{{\gamma}_d}{\gamma^{(c)}}=\frac{1}{12}$ would yield $n_S \simeq 0.96$. According to eq.~\eqref{gammaovergammac}, this result would obtain for $\alpha=2$ and $p=4$, \emph{i.e.} for an $NS$ five--brane wrapped around a small defect in the extra dimensions, which would make it effectively look like a four--dimensional extended object.  This would be naturally available only for the $SO(16) \times SO(16)$ heterotic string of~\cite{so16xso16}, which motivates us to take a closer look at dualities for these non--supersymmetric strings, starting from the original work in~\cite{dienes}.

\section{Linear--Dilaton Cosmologies} \label{sec:lin_dil}

In the preceding section we left out the special case $\widetilde{\gamma}=\gamma_\ell$, in which the solutions in eqs.~\eqref{sol_cosmo_d} become singular. As explained in Appendix~\ref{sec:app4}, the solutions in cosmic time are not singular, and the problem arises merely from the definition of conformal time at the transition between accelerated and decelerated cosmologies. The correct procedure to deal with this issue is to look for a different class of solutions of the system~\eqref{cosmo_eqs_d}, with $\phi'$ and $\Omega'$ constant. These describe the linear--dilaton cosmology of~\cite{lindil_2} and read
\beq
\Omega \ = \  a\, \eta \ , \qquad
\phi  \ = \  - \ \frac{a}{2} \, (d-2)\, \eta \ , \label{lindilsol}
\eeq
where the parameter $\alpha$ is related to the overall strength of the potential in eq.~\eqref{lagrangian_bsb_D}
according to
\beq
e^{2\Omega}\,V_0 \ = \ (d-2)^2\, a^2 \ . \label{lindilpot}
\eeq

As a result the equations for scalar and tensor perturbations become identical, and reduce to the simple form
\beq
A'' \ + \ A'\,(d-2)\,a \ + \ {\mathbf{k}^2}\, A \ = \ 0 \ .
\eeq
whose solutions
clearly decay in an expanding Universe for all values of ${\mathbf{k}}$.

This linear--dilaton background is an exact solution of tree--level String Theory to all orders in $\alpha'$~\cite{lindil_2}, with a strong--coupling region in the infinite past, for large negative values of $\eta$, at the initial singularity of the Einstein--frame metric. It is conceivable that string corrections involving higher powers of $g_s$ could cure the problem, and the fact that these solutions correspond to string--frame Minkowski metrics appears encouraging to this effect.
Once more, the behavior of cosmological solutions of String Theory in the presence of broken supersymmetry appears encouraging.

\section{Conclusions} \label{sec:conclusions}

In this paper we have explored the perturbative stability of four classes of backgrounds that arise when the low--energy effective action of String Theory is supplemented by the dilaton potentials that are by the breaking of Supersymmetry. Our results are as follows.
\begin{enumerate}
\item In the $AdS_3 \times S^7$ vacua of the $USp(32)$ and $U(32)$ orientifolds of~\cite{ms17}
    \begin{enumerate}
    \item \emph{no instabilities are present in $AdS$ tensor and vector modes}, and also in the (anti)sym\-metric tensors in internal space $h_{ij}$ and $b_{ij}$;
    \item \emph{instabilities are present in scalar modes} that draw their origin from the dilaton, the metric tensor and the two--form, as explained in Section~\ref{sec:perturbations37_gen}. These expand into totally symmetric and traceless $SO(8)$ tensors, and \emph{the problem is present for angular momenta corresponding to the three values $\ell=2,3,4$}, in agreement with~\cite{gubsermitra}, but have also explored projections of the internal sphere that remove them, perhaps at the price of fixing some sub--varieties. However, even leaving aside the issue of sub--varieties, one ought to investigate vacuum bubbles, which typically occur in Kaluza--Klein theory~\cite{wittenkk}, and vacuum bubbles were indeed shown to destabilize vacua with similar projections in~\cite{hop}. We leave this problem for future work;
    \item \emph{wide stability regions} exist, however, close to the values of ${V_0}$, ${V_0'}$ and $V_0''$ for the (projective--)disk potential of eq.~\eqref{pot32}. These regions, where $V_0<0$, could be relevant if corrections are taken into account.
    \end{enumerate}
  \item The $AdS_7 \times S^3$ vacua of the $SO(16) \times SO(16)$ heterotic string display a similar behavior, but:
      \begin{enumerate}
    \item\emph{ instabilities} are present in scalar modes only for $\ell=1$, but if the internal sphere is replaced with an orbifold determined by an antipodal $\mathbb{Z}_2$ projection, the unstable modes of the tree--level potential disappear. We leave the problem of vacuum bubbles for future work also in this case;
    \item wide stability regions exist again, close to the values of ${V_0}$, ${V_0'}$ and $V_0''$ for the torus--potential of eq.~\eqref{pot52}. These regions, where $V_0<0$, could again be relevant if corrections are taken into account;
    \end{enumerate}
  \item \emph{the vacua of~\cite{dm_9Dsolution} with nine--dimensional Poincar\'e symmetry are perturbatively stable}, for both the orientifold and heterotic models. Or, more precisely, they are perturbatively stable solutions of Einstein--dilaton systems, since from the vantage point of String Theory they include regions of strong coupling;
  \item the \emph{non--homogeneous perturbations of the cosmological solution of~\cite{dm_9Dsolution} are stable for large times, while a logarithmic growth, which points to a breaking of isotropy, is present for the homogeneous tensor mode}. We could not resist to notice that broken supersymmetry in String Theory, with its steep runaway lowest--order potentials, points in this fashion to an awaited origin for compactification to $d< 10$. Interestingly, however, this would not select the final value of $d$, at variance with the indications of~\cite{anagnostopoulos};
\item more general exponential potentials, whose cosmologies were first discussed in~\cite{russo} and are central to the ``climbing scalar'' picture of~\cite{climbing}, display a sharp change of behavior, in ten dimensions, when the parameter $\gamma_E$ of Section~\ref{sec:general_exponential} reaches the value $\frac{3}{2}$, where the climbing behavior sets it.
    \emph{Below this ``critical'' value $\gamma^{(c)}$  the cosmological solutions approach at large times the $d$--dimensional counterparts of the Lucchin--Matarrese attractor of~\cite{lm}, and then perturbations do not grow};~\footnote{As we have explained in Section~\ref{sec:general_exponential}, in~\cite{climbing} we used dimension--dependent normalizations for the scalar field and the parametric time of~\cite{dm_9Dsolution}, in such a way that the onset of the climbing behavior always took place at $\gamma=1$. The conversions to the present notation are collected in eq.~\eqref{redefinitions}.}
\item the special case $\widetilde{\gamma}=\gamma_\ell$ must be treated separately in conformal time, but all perturbations are stable and have a simple behavior. This special value also corresponds to the linear--dilaton cosmologies of~\cite{lindil_2}, which arise above the critical dimension of String Theory and are exact tree--level solutions to all orders in $\alpha'$ where supersymmetry is broken at the sphere level.

\end{enumerate}
A fair summary of the indications of this work for String Theory is that, while problems emerge in non--supersymmetric flux vacua, the situation appears more natural and promising in evolving cosmological settings. Taking seriously the hint that we gathered in Section~\ref{sec:tensor_climbing}, there would be even clues for the origin of compactification. This all brings to our mind the time--honored case of Newtonian interactions, where static instability can leave way to dynamical stability, with a wealth of manifestations that underlie the Universe up to the largest scales that we have explored to date. Or, for that matter, the case of static Coulomb instabilities, with all they contributed to our understanding of Nature via Quantum Mechanics. Time will tell whether these insights, together with many others that are starting to surface~\cite{weak_gravity}, will converge into a main hidden lesson from String Theory.

There are some possible extensions of this work, some of which are under investigation:
\begin{enumerate}
\item the general solutions in~\cite{ms17} also included internal gauge fields. These complicate the vacua, and seem not to improve matters for stability, but it may be interesting to complete the analysis;

\item bubble formation can introduce non--perturbative instabilities when internal orbifold projections remove perturbative ones, as in~\cite{wittenkk}. Taking a closer look at this problem would be particularly relevant for the heterotic $AdS_7 \times S^3$ vacua, where a simple $\mathbb{Z}_2$ internal projection can remove all unstable modes;

\item in lower dimensions, one could combine Brane Supersymmetry Breaking with Scherk--Schwarz reductions~\cite{scherkschwarz}, along the lines of what was done for closed strings in~\cite{closed_ss} and for open strings in~\cite{open_ss}, and it would be interesting to explore the stability of corresponding vacua;

\item it would be interesting to see how dilaton potentials deform the $D_p$-branes of the models at stake. Without taking these deformations into account, their world--sheet description was presented in~\cite{dms};

 \item finally, the duality between heterotic and orientifold models becomes perturbative for $d<6$, and has already displayed some interesting consequences in the supersymmetric case~\cite{three_generations}. The considerations at the end of Section~\ref{sec:general_exponential}, and the apparently richer options allowed for the heterotic $SO(16) \times SO(16)$ model, motivate a closer look at the string duality web with broken supersymmetry in $d < 10$, relying on the steps made in~\cite{dienes}.
\end{enumerate}
\begin{appendices}
\section{An Equation Relevant for Scalar Perturbations} \label{sec:app1}
Let us consider an equation of the form
\beq
\lambda_{\mu\nu}\, A \ + \ \nabla_\mu\,\nabla_\nu\,B \ = \ 0 \ , \label{mixed_eq}
\eeq
or
\beq
\gamma_{ij}\, A \ + \ \nabla_i\,\nabla_j\,B \ = \ 0 \ ,
\eeq
which appeared recurrently, in the analysis of scalar perturbations, in Sections~\ref{sec:scalar_pert_37}, \ref{sec:scalar_pert_73} and \ref{sec:scalar9d}. Referring for definiteness to the first form, we would like to prove that this type of equation implies that $A$ and $B$ must both vanish.

To begin with, one can take a trace, and if the $AdS$ space is of dimension $n_1$ this gives
\beq
n_1\, A \ + \ \Box\,B \ = \ 0 \ . \label{trace_eq}
\eeq
Next one can take the divergence, obtaining finally
\beq
A\ + \Box\,B \ + \  \frac{1-n_1}{R_{AdS}^2} \ B \ = \ 0 \ .
\eeq
Subtracting from this eq.~\eqref{trace_eq} one now finds
\beq
A \ + \ \frac{1}{R_{AdS}^2} \ B \ = \ 0 \ ,
\eeq
and consequently the original eq.~\eqref{mixed_eq} can be recast in the form
\beq
\nabla_\mu\,\nabla_\nu\,B \ = \  \frac{1}{R_{AdS}^2}\, \lambda_{\mu\nu} \ B \ .
\eeq
If one now assumes that $B \neq 0$, letting $C = \log B$, this result can be turned into
\beq
\nabla_\mu\,\nabla_\nu\,C + \nabla_\mu\,C \, \nabla_\nu\, C \ = \ \frac{1}{R_{AdS}^2}\, \lambda_{\mu\nu} \ .
\eeq
Redefining the background metric by a coordinate transformation, one can remove the first term, but the resulting equation
\beq
\nabla_\mu\,C \, \nabla_\nu\, C \ = \ \frac{1}{R_{AdS}^2}\, \lambda_{\mu\nu} \ .
\eeq
is inconsistent, since the \emph{l.h.s.} is clearly a matrix of lower rank. Hence $B=0$ and therefore, a fortiori, $A=0$.
\section{On Tensor Spherical Harmonics in $n$ Dimensions} \label{sec:app2}
In this Appendix we review some results that were needed in our analysis in Sections~\ref{sec:perturbations37_gen} and \ref{sec:perturbations73_gen}, starting from an ambient Euclidean space~\cite{book_spherical}. The results agree with the constructions presented in~\cite{tensor_harmonics}.
If $Y^1,\dots Y^{n+1}$ are Cartesian coordinates of $R^{n+1}$,
so that the unit sphere $S^{n}$ is described by the quadratic constraint
\beq
\delta_{IJ}\,Y^I\, Y^J\ =\ 1 \ ,
\eeq
the scalar spherical harmonics on $S^{n}$ can be conveniently constructed starting from harmonic polynomials of degree $\ell$ in the embedding Euclidean space $R^{n+1}$. A harmonic polynomial of degree $\ell$
\beq
H_{(n)}^{\ell}(Y)\ =\ \a_{I_1\dots I_\ell}\, Y^{I_1}\dots Y^{I_\ell}
\eeq
is determined by a \emph{totally symmetric and traceless} tensor $\a_{I_1\dots I_\ell}$ of rank $\ell$, as can be clearly seen applying to it the Laplace operator
\beq
\nabla^2 \ = \ \sum_{I=1}^{n+1}\  \frac{\partial^2}{\partial\,Y_I^{\,2}}
\eeq
in Cartesian coordinates.

The scalar spherical harmonics ${\cal Y}_{(n)}^{I_1\dots I_\ell}$ are defined restricting the $H_{(n)}^{\ell}(Y)$ to the unit sphere $S^n$, or equivalently as
\beq
H_{(n)}^{\ell}(Y) \ = \ r^\ell\, \a_{I_1\dots I_\ell}\,{\cal Y}_{(n)}^{I_1\dots I_\ell}\left( y^i \right) \ = \ r^\ell\, \a_{I_1\dots I_\ell}\, \hat{Y}^{I_1}\dots \hat{Y}^{I_\ell}\ ,
\eeq
in terms of the radial coordinate
\beq
r^2\ =\ \delta_{IJ}\, Y^I\, Y^J \ ,
\eeq
and of the ${\hat Y}^I$, where
\beq
Y^I= r\,\hat{Y}^I(y^i) \ .  \label{cart_to_polar}
\eeq
Here $i=1,\ldots , n$, and the $y^i$ are a set of coordinates on $S^n$. As a result, the metric takes the form
\beq
ds^2 \ = \ \delta_{IJ}\,dY^I\, dY^J \ = \ dr^2 \ + \ r^2 \ ds^2_{S^n} \ , \label{spherical_decomp}
\eeq
and the Laplacian on scalars decomposes according to
\beq
0 \ = \ \left(\nabla^2\right)_{R^{n+1}} H_{(n)}^{\ell}(Y) \ = \ \frac{1}{r^n} \ \frac{\partial}{\partial\, r} \left( r^n \, \frac{\partial\, H_{(n)}^{\ell}(Y)}{\partial r} \right) \ + \ \frac{1}{r^2}\ \left(\nabla^2\right)_{S^n} H_{(n)}^{\ell}(Y) \ ,
\eeq
where
\beq
\frac{\partial H_{(n)}^{\ell}(Y)}{\partial r} \ = \ \frac{\ell}{r} \ H_{(n)}^{\ell}(Y) \
\eeq
for the homogeneous polynomials $H_{(n)}^{\ell}(Y)$. All in all
\beq
\left(\nabla^2\right)_{S^n} {\cal Y}_{(n)}^{I_1\dots I_\ell} \ = \ - \ \ell(\ell+n-1) \ {\cal Y}_{(n)}^{I_1\dots I_\ell} \ ,
\eeq
and the degeneracy of the scalar spherical harmonics for any given $\ell$ is the number of independent components of a corresponding totally symmetric traceless tensor, \emph{i.e.}
\beq
\frac{(n+2 \ell -1)(n + \ell -2)!}{\ell!(n - 1)!}  \ .
\eeq

In discussing more general tensor harmonics, it is convenient to notice that, in the coordinate system of eq.~\eqref{spherical_decomp}, the non--vanishing Christoffel symbols ${\widetilde \Gamma}_{IJ}^K$ for the ambient Euclidean space read
\beq
{\widetilde \Gamma}_{ij}^r \ = \ - \ r\, g_{ij} \ , \quad {\widetilde \Gamma}_{jr}^i = \  \frac{1}{r}\, \delta_i^{\,j} \ , \quad {\widetilde \Gamma}_{ij}^k \ = \  \Gamma_{ij}^k \ . \label{christoffel_total}
\eeq
The labels $i,j,k$ refer, as above, to the $n$--sphere, whose Christoffel symbols are denoted by $\Gamma_{ij}^k$.

The construction extends nicely to tensor spherical harmonics, which can be defined starting from generalized harmonic polynomials, with one proviso. The relation \eqref{cart_to_polar} and its differentials imply that the actual spherical components of tensors carry additional factors of $r$, one for each covariant tensor index, with respect to those naively inherited from the Cartesian coordinates of the Euclidean ambient space, as we shall now see in detail.
To begin with, vector spherical harmonics obtain starting from one--forms in ambient space, built from harmonic polynomials of the type
\beq
H_{(n);\,J}^{\ell}(Y)\ =\ \a_{I_1 \dots I_\ell\,;\,J}\, Y^{I_1}\dots Y^{I_\ell}\ ,
\eeq
where the coefficients $\a_{I_1 \dots I_\ell\,,\,J}$ are \emph{totally symmetric and traceless in any pair of $I$--indices}. They are also subject to the condition
\beq
Y^J\, H_{(n);\,J}^{\ell}(Y) \ = \ 0 \ ,
\eeq
since the radial component, which does not pertain to the sphere $S^n$, ought to vanish. This implies that the total symmetrization of the coefficients vanishes identically,
\beq
\a_{\left(I_1\dots I_\ell\,;\,I\right)} \ = \ 0 \ ,
\eeq
and on account of the total symmetry in the $I$--labels is $H_{n\,,\,J}^{\ell}(Y)$ is thus \emph{transverse} in the ambient space:
\beq
\partial^J \, H_{(n);\,J}^{\ell}(Y) \ =  \ 0 \ .
\eeq
Moreover, any Cartesian vector $V$ such that $V_I\, Y^I =0$ couples with differentials according to the general rule inherited from eq.~\eqref{cart_to_polar},
\beq
V_I\, dY^I \ = \ V_I \, r\, d {\hat Y}^ I \ ,
\eeq
so that the actual sphere components, which are associated to the $d {\hat Y}^ I$, include an additional power of $r$, and the vector spherical harmonics ${\cal Y}_{(n)\, i}^{I_1\dots I_\ell;J}$ are thus obtained from
\beq
r^{\ell+1}\,{\cal Y}_{(n)\, i}^{I_1\dots I_\ell;J} \, \alpha_{I_1 \ldots I_\ell;J}\, \, d {y}^i \ = \ r\,H_{(n);\,J}^{\ell}(Y)\, d {\hat Y}^ J \ .
\eeq
As a result,
\beq
\nabla_r\, \nabla_r \, \left(r\,H_{(n);I}^{\ell}(Y)\right) \ = \ \left(\frac{\partial}{\partial r} \ - \ \frac{1}{r} \right)^2 \left(r\,H_{(n);J}^{\ell}(Y)\right) \ = \ \frac{\ell(\ell -1)}{r} \ H_{(n);I}^{\ell}(Y) \ , \label{eq_vector_1}
\eeq
while the remaining contributions to the Laplacian give
\beq
\frac{1}{r^2} \ \left(\nabla^2\right)_{S^n} \left(r\,H_{(n);I}^{\ell}(Y)\right) \ + \ \frac{n(\ell+1) - n -1}{r}\ \left(r\,H_{(n);I}^{\ell}(Y)\right) \ , \label{eq_vector_2}
\eeq
taking into account the Christoffel symbols in eq.~\eqref{christoffel_total}. Since the total Euclidean Laplacian vanishes by construction, adding eqs.~\eqref{eq_vector_1} and \eqref{eq_vector_2} gives finally
\beq
\left(\nabla^2\right)_{S^n}\,{\cal Y}_{(n)\, i}^{I_1\dots I_\ell;J} \ = \ - \ \Big[ \ell(\ell+n-1) \ - \ 1 \Big] \,{\cal Y}_{(n)\, i}^{I_1\dots I_\ell;J} \ ,
\eeq
with $\ell \geq 1$.

In a similar fashion, the spherical harmonics ${\cal Y}_{(n)\, i_1\ldots i_p}^{I_1\dots I_\ell;J_1 \ldots J_p}$ corresponding to generic higher--rank transverse tensors, which are also traceless in any pair of symmetric $I$--indices, can be described starting from harmonic polynomials of the type $H_{(n);I_1\ldots I_p}^{\ell}(Y)$, and satisfy
\beq
\left(\nabla^2\right)_{S^n}\,{\cal Y}_{(n)\, i_1\ldots i_p}^{I_1\dots I_\ell;J_1 \ldots J_p} \ = \ - \ \Big[ \ell(\ell+n-1) \ - \ p \Big] \,{\cal Y}_{(n)\, i_1\ldots i_p}^{I_1\dots I_\ell;J_1 \ldots J_p} \ ,
\eeq
with $\ell \geq p$.

In Young tableaux language, the scalar harmonics correspond to \emph{traceless } single--row diagrams of the type
\beq
\ytableausetup
{mathmode, boxsize=1.9em}
\begin{ytableau}
I_1 & I_2 & \none[\dots]
& I_l
\end{ytableau} \ \ \ ,
\eeq
while the independent vectors associated to vector harmonics correspond to two--row \emph{traceless} hooked diagrams of the type
\beq
\ytableausetup
{mathmode, boxsize=1.9em}
\begin{ytableau}
I_1 & I_2 & \none[\dots]
& I_l \\
J
\end{ytableau} \ \ \ ,
\eeq
as we have explained.
In a similar fashion, the independent tensor metric perturbations in the internal space correspond to \emph{traceless} diagrams of the type
\beq
\ytableausetup
{mathmode, boxsize=1.9em}
\begin{ytableau}
I_1 & I_2 & \none[\dots]
& I_l \\
J_1 & J_2
\end{ytableau} \ \ \ ,
\eeq
while the independent perturbations associated to a $(p+1)$--form gauge field in the internal space correspond, in general, to multi--row diagrams of the type
\beq
\ytableausetup
{mathmode, boxsize=1.9em}
\begin{ytableau}
I_1 & I_2 & \none[\dots]
& I_l \\
J_1 \\
\none[\vdots] \\
J_{p+1}
\end{ytableau} \ \ \  .
\eeq

The degeneracies of these representations can be related to the corresponding Young tableaux, for instance, as in~\cite{book_spherical_ma}.
The structure of the various types of harmonics, which are genuinely different for large enough values of $n$, reflects nicely the generic absence of mixings between different classes of perturbations.
\section{Some Breitenlohner--Freedman Bounds} \label{sec:app3}

In this Appendix we collect some Breitenlohner--Freedman (BF) bounds~\cite{BF} that play a role in the preceding sections. To this end, it is convenient to work in conformally flat Poincar\'e coordinates, so that the $AdS_d$ metric takes the form
\beq
ds^2 \ = \ R_{AdS}^2\ g_{MN}\, d x^M\,dx^N \ = \ R_{AdS}^2\ \frac{dz^2 \ + \ \eta_{\mu\nu}\,dx^\mu\,dx^\nu}{z^2}  \ ,
\eeq
where $(\mu,\nu=0,..,d-2)$. The non--vanishing Christoffel connections are then
\beq
\Gamma_{zz}^z \ = \ - \frac{1}{z} \ , \quad \Gamma_{\nu z}^\mu \ = \ -\ \frac{1}{z} \ \delta_\nu^\mu \ , \quad \Gamma_{\mu \nu}^z \ = \  \frac{1}{z} \ \eta_{\mu\nu} \ .
\eeq

In this coordinate system the scalar Klein--Gordon equation
\beq
g^{MN} \, \nabla_M\, \nabla_N \,\Phi  \ - \ \left( M\, R_{AdS}\right)^2\, \Phi \ = \ 0
\eeq
takes the form
\beq
f'' \ + \ \frac{2-d}{z}\, f' \ - \ \left(k^2 \,+\,\frac{\left( M\, R_{AdS}\right)^2}{z^2}\right) f \ = \ 0 \ ,
\eeq
where ``primes'' denote $z$--derivatives and we are focussing on plane waves of the type
\beq
\phi(x,z) \ = \ e^{i k \cdot x} \, f(z) \ .
\eeq
Letting now
\beq
f(z) \ = \ \Psi(z) \, z^{\frac{d}{2} - 1}
\eeq
reduces the field equation to the Schr\"odinger--like form
\beq
\left[ - \ \frac{d^2}{dz^2} \ + \ \frac{4\,\left( M\, R_{AdS}\right)^2+d\left( d \, - \, 2\right)}{4\,z^2}\right] \Psi \ = \ - \ k^2\, \Psi \ ,
\eeq
so that the operator acting on $\phi$ can be presented as in Section~\ref{sec:pert_9D_vacuum},
\beq
- k^2\,\Psi \ = \ {\cal A}^\dagger\, {\cal A}\, \Psi \ ,
\eeq
where now
\beq
{\cal A} \ = \ - \ \frac{d}{dz} \ + \ \frac{a}{z} \ , \quad {\cal A}^\dagger \ = \  \frac{d}{dz} \ + \ \frac{a}{z} \ ,
\eeq
and
\beq
\left( a \ + \ \frac{1}{2} \right)^2 \ = \ \left( M\, R_{AdS}\right)^2 \ + \ \frac{(d-1)^2}{4} \ .
\eeq
Requiring that $-k^2>0$, \emph{i.e.} the absence of tachyonic excitations, and thus of modes potentially growing in time, in the Minkowski sections at constant $z$, translates into the condition that $a$ be real, and hence into the BF bound
\beq
\left( M\, R_{AdS}\right)^2 \ + \ \frac{(d-1)^2}{4} \ \geq \ 0 \ .
\eeq

One can treat the case of a massive vector in a similar fashion, starting from
\beq
{\cal S} \,=\, \int d^d x \, \sqrt{-g} \left[ - \ \frac{1}{4} \ F_{M_1 M_2}\,F_{N_1 N_2}\, g^{M_1N_1} \, g^{M_2 N_2}
\,-\, \frac{M^2}{2} \ A_{M}\,A_{N}\, g^{M N} \right] \ .
\eeq
To begin with, let us recall that the massive Proca equation implies the divergence--free condition, which in Poincar\'e coordinates reads
\beq
\partial \cdot A \ + \ \frac{2-d}{z} \ A_z \ + \ \partial_z\,A_z \ = \ 0 \ , \label{div_constraint}
\eeq
and that, after letting
\beq
A_M \ = \ e^{\,i \, k \cdot x}\, f_M(z) \ ,
\eeq
the resulting dynamical equation,
\beq
R_{AdS}^2\,\Box\, A_N \ + \ \left[d\,-\,1\,-\, (M \, R_{AdS})^2 \right] A_N \ = \ 0
\eeq
translates into
\bea
\partial_z^2\, A_z &+& \frac{2-d}{z}\, \partial_z\,A_z \ + \ \left[ \,-\,k^2\ + \ \frac{d-2-(M \, R_{AdS})^2}{z^2} \right] A_z \ = \ 0 \ , \nonumber \\
\partial_z^2\, A_\mu &+& \frac{4-d}{z}\, \partial_z\,A_\mu \ - \ \left[ k^2\ + \ \frac{(M \, R_{AdS})^2}{z^2} \right] A_\mu \ = \ \frac{2}{z}\ \partial_\mu\, A_z \ . \label{vector_bound_eqs}
\eea

Changing variable as was done for the scalar, one can see that the first of these leads to the condition
\beq
\left( a \ + \ \frac{1}{2} \right)^2 \ = \ \left( M\, R_{AdS}\right)^2 \ + \ \frac{(d-3)^2}{4} \ ,
\eeq
from which one can infer the bound in $AdS_d$,
\beq
\left( M\, R_{AdS}\right)^2 \ + \ \frac{(d-3)^2}{4} \ \geq \ 0 \ . \label{BF_vector}
\eeq
Let us stress that this bound refers to the mass term in the Lagrangian, since we have subtracted the contribution arising from commutators, which is also present in the massless case.
The second of eqs.~\eqref{vector_bound_eqs} is apparently more complicated, since it contains $A_z$ as a source. However, one can now distinguish in it the longitudinal and transverse portions of $A_\mu$. The former can be related to $A_z$ via eq.~\eqref{div_constraint}, and one is lead again, for it, to the first of eqs.~\eqref{vector_bound_eqs}. The latter satisfies the same equation as the scalar field, but for the replacement of $d$ with $d-2$. All in all, one is thus led again to the BF bound \eqref{BF_vector}.

One can also study rather simply the general case of
form potentials $B_{p+1}$, starting from their action principles
\bea
{\cal S} &=& \int d^d x \, \sqrt{-g} \left[ - \ \frac{1}{2(p+2)!} \ H_{M_1 \ldots M_{p+2}}\,H_{N_1 \ldots N_{p+2}}\, g^{M_1N_1} \cdots  g^{M_{p+2}N_{p+2}} \right. \nonumber \\
&-& \left. \frac{M^2}{2(p+1)!} \ B_{M_1 \ldots M_{p+1}}\,B_{N_1 \ldots N_{p+1}}\, g^{M_1N_1} \cdots  g^{M_{p+1}N_{p+1}} \right] \ ,
\eea
and for a transverse $B$ the equation of motion can be recast in the form
\beq
\Box\,B_{M_1 \ldots M_{p+1}} \ + \ \left[ \frac{(p+1)(d-p-1)}{R_{AdS}^2} \ - \ M^2 \right]\, B_{M_1 \ldots M_{p+1}} \ = \ 0 \ . \label{massive_pform}
\eeq
Moreover, in Poincar\'e coordinates the action takes the form
\beq
{\cal S} \ = \ \int d^d x \, \frac{1}{z^d} \left[ - \ \frac{1}{2(p+2)!} \ z^{2(p+2)} H^2 \ - \ z^{2(p+1)} \, \frac{M^2}{2(p+1)!} \ B^2 \right] \ ,
\eeq
where indices are raised and lowered with a $d$--dimensional flat metric.
Extending the preceding discussion, one can thus conclude that the BF bounds in $AdS_d$ on the mass $M$ for $(p+1)$--form gauge fields are
\beq
\left( M\, R_{AdS}\right)^2 \ + \ \frac{(d-3-2p)^2}{4} \ \geq \ 0 \ , \label{BF_form}
\eeq
a result that applies insofar $d>p+2$. Notice that this relation, which refers again to the mass term in the Lagrangian, is invariant under the ``massive duality'' (see, for instance, \cite{fs15}) between $(p+1)$--form and $(d-p-2)$--form fields.

\section{The Attractor Solutions in Cosmic and Conformal Time} \label{sec:app4}

It is instructive to take a look at the attractor solutions of Section~\ref{sec:general_exponential} in cosmic time, which seem singular for $\widetilde{\gamma} = \gamma_\ell$ in conformal time. In Section~\ref{sec:general_exponential} we were led to eqs.~\eqref{sol_cosmo_d}, and the first of them, with a suitable choice of integration constant, determines indeed the metric
\beq
ds^2 \ = \ \left(\frac{\eta}{\lambda+1}\right)^{2\,\lambda} \Big[ - d \eta^2 \ + \ d{\mathbf{x}} \cdot d{\mathbf{x}} \Big] \ ,
\eeq
where
\beq
\lambda \ = \ \frac{1}{\left[\left(\frac{\widetilde{\gamma}}{\gamma_\ell}\right)^2 - \, 1\right]}
 \ .
\eeq

The corresponding expression in cosmic time reads
\beq
ds^2 \ = \ -\ dt^2 \ +  \ t^{\frac{2\,\lambda}{\lambda+1}} \ d {\mathbf x} \cdot d {\mathbf x} \ , \label{cosmic_cosmo_d}
\eeq
and is regular as $\lambda \to \infty$, a limit where the Universe evolves in time with no acceleration. Here
\beq
t \ = \ \left( \frac{\eta}{\lambda+1}\right)^{\lambda+1} \ ,
\eeq
and moreover, expressing the third of eqs.~\eqref{sol_cosmo_d} in terms of cosmic time also leads to a regular expression,
\beq
e^{2\Omega}\, V \ = \ \frac{(d-2)^3}{16} \ \frac{\Big[{\gamma^{(c)\,2}} \ - \ \widetilde{\gamma}^2\Big]}{t^{\,\frac{2}{\lambda+1}}}\ \left(\frac{\gamma_\ell}{{\widetilde \gamma}}\right)^4 \ ,
\eeq
which tends to a constant as $\lambda \to \infty$, consistently with the discussion in Section~\ref{sec:lin_dil}.

\end{appendices}

\vskip 12pt
\noindent {\large \bf Acknowledgements}

\noindent J.M. is grateful to Scuola Normale Superiore and A.S. is grateful to the CPhT--\'Ecole Polytechnique, APC--U. Paris VII and the CERN Theory Department for the kind hospitality while this work was in progress. I.~B. and A.~S. were supported in part by Scuola Normale Superiore and by INFN (IS CSN4-GSS-PI). We are grateful to I.~Antoniadis and H. Partouche for correspondence, to N.~Kitazawa for useful suggestions and especially to E.~Dudas, A.~Malchiodi and A.~Tomasiello for stimulating discussions and suggestions.
\newpage
\setcounter{equation}{0}

\end{document}